\documentclass[twocolumn]{aastex631}

\usepackage{graphicx}	
\usepackage{amsmath}	
\usepackage{graphicx}	
\usepackage{color}
  
\usepackage[figuresright]{rotating}
\usepackage{enumerate}
\usepackage{verbatim}
\usepackage{url}
\usepackage{ulem}
\usepackage{float}
\hypersetup{citecolor=blue,urlcolor=blue}
\usepackage{hyperref}
\usepackage{verbatim}
\usepackage{url}
\usepackage{float} 
\usepackage[T1]{fontenc}
\usepackage[utf8]{inputenc}

\usepackage{mathrsfs}

\begin{document}

\title{Component-dependent Periodic Emission Variations of
PSR J1857+0943 Observed with FAST}


\author[0000-0002-2060-5539]{{Shi-jun Dang}}
\affiliation{Guizhou Radio Astronomical Observatory, Guiyang 550001, People's Republic of China\\}
\affiliation{School of Physics and Electronic Science, Guizhou Normal University, Guiyang 550001, People's Republic of China\\}

\author{Ji-guang Lu}
\email{lujig@nao.cas.cn(J. G. Lu)}
\affiliation{Guizhou Radio Astronomical Observatory, Guiyang 550001, People's Republic of China\\}
\affiliation{CAS Key Laboratory of FAST, NAOC, Chinese Academy of Sciences, Beijing 100101, People's Republic of China\\}
\affiliation{National Astronomical Observatories, Chinese Academy of Sciences, Beijing 100101, People’s Republic of China}

\author{Peng Jiang}
\email{pjiang@nao.cas.cn(P. Jiang)}
\affiliation{Guizhou Radio Astronomical Observatory, Guiyang 550001, People's Republic of China\\}
\affiliation{CAS Key Laboratory of FAST, NAOC, Chinese Academy of Sciences, Beijing 100101, People's Republic of China\\}
\affiliation{National Astronomical Observatories, Chinese Academy of Sciences, Beijing 100101, People’s Republic of China}

\author[0000-0001-9986-9360]{Yu-lan Liu}
\affiliation{Guizhou Radio Astronomical Observatory, Guiyang 550001, People's Republic of China\\}
\affiliation{CAS Key Laboratory of FAST, NAOC, Chinese Academy of Sciences, Beijing 100101, People's Republic of China\\}
\affiliation{National Astronomical Observatories, Chinese Academy of Sciences, Beijing 100101, People’s Republic of China}

\author{Jin-tao Xie}
\affiliation{School of Computer Science and Engineering, Sichuan University of Science and Engineering, Yibin 644000, China}

\author[0000-0002-5815-6548]{Habtamu Menberu Tedila}
\affiliation{CAS Key Laboratory of FAST, NAOC, Chinese Academy of Sciences, Beijing 100101, People's Republic of China\\}
\affiliation{National Astronomical Observatories, Chinese Academy of Sciences, Beijing 100101, People’s Republic of China}
\affiliation{Arba Minch University, Arba Minch 21, Ethiopia}

\author[0000-0002-9786-8548]{Fei-fei Kou}
\affiliation{XinJiang Astronomical Observatory, CAS, Urumqi, Xinjiang 830011, People's Republic of China\\}

\author{Jian-ping Yuan}
\affiliation{XinJiang Astronomical Observatory, CAS, Urumqi, Xinjiang 830011, People's Republic of China\\}

\author{Zhi-gang Wen}
\affiliation{XinJiang Astronomical Observatory, CAS, Urumqi, Xinjiang 830011, People's Republic of China\\}

\author[0000-0003-4498-6070]{Shuang-qiang Wang}
\affiliation{XinJiang Astronomical Observatory, CAS, Urumqi, Xinjiang 830011, People's Republic of China\\}

\author{Lun-hua Shang}
\affiliation{School of Physics and Electronic Science, Guizhou Normal University, Guiyang 550001, People's Republic of China\\}

\author[0009-0005-8224-0677]{Zu-rong Zhou}
\affiliation{National Time Service Center, Chinese Academy of Sciences, Xi’an 710600, China}

\author{Wen-ming Yan}
\affiliation{XinJiang Astronomical Observatory, CAS, Urumqi, Xinjiang 830011, People's Republic of China\\}

\author{Qi-jun Zhi}
\affiliation{College of Physics, Guizhou University, Guiyang, Guizhou 550001, People's Republic of China\\}
\affiliation{School of Physics and Electronic Science, Guizhou Normal University, Guiyang 550001, People's Republic of China\\}

\author{Na Wang}
\affiliation{XinJiang Astronomical Observatory, CAS, Urumqi, Xinjiang 830011, People's Republic of China\\}


\begin{abstract}
Single-pulse studies of millisecond pulsars provide an important way to probe the plasma processes responsible for their radio emission, but such studies remain rare because of the weak emission and short rotation periods of these objects. Using the high sensitivity of the Five-hundred-meter Aperture Spherical radio Telescope (FAST), we investigate the pulse-to-pulse variability of PSR J1857+0943 and report the first detection of distinct fluctuation behavior across its main pulse and interpulse. The integrated main pulse and interpulse each show multiple emission components with markedly different variability properties. Pulse longitude resolved fluctuation analysis reveals quasi-periodic intensity modulation in both pulse regions, with the interpulse showing a characteristic modulation timescale approximately twice that observed in the trailing part of the main pulse. A longitude-resolved cross-correlation analysis further reveals a weak anti-correlation between the trailing main-pulse emission and the interpulse at a lag of one rotation period, suggesting that these widely separated pulse-longitude ranges may be influenced by a common magnetospheric process. The polarization properties of the single pulses in the same longitude range also differ from those of the adjacent main-pulse components, supporting the presence of component-dependent emission behavior. In addition, we measure a one-hour jitter noise of $\sigma_{\rm J,1h}=73.2\pm1.5$ ns at 1.25 GHz, consistent with previous measurements. These results demonstrate that PSR J1857+0943 exhibits complex single-pulse variability and provide new observational constraints on the magnetospheric structure of interpulse millisecond pulsars.
\end{abstract}


\section{Introduction} \label{sec:intro}

Millisecond pulsars (MSPs) are known for their exceptionally stable periods, making them amongst the best natural timekeepers in the universe. Their timing stability makes them valuable for probing the dense-matter physics of neutron stars \citep{2016ARA&A..54..401O,2020ApJ...892....4D}, testing theories of gravity \citep{2014Natur.505..520R,2015ApJ...809...41Z,2023ApJ...958L..17W}, and detecting the nanohertz gravitational-wave background \citep[e.g.,][]{1992RSPTA.341..117T,2020PASA...37...20K,2022MNRAS.509.5538C,2023RAA....23g5024X,2023ApJ...951L...9A,2023A&A...678A..48E}. Observations of MSPs also provide information about the structure and turbulence of the interstellar medium \citep{2015ApJ...808..113C,2023SCPMA..6619512L,2023SCPMA..6699511Z} and the Galactic magnetic field \citep{2018ApJS..234...11H}. Long-term pulsar timing observations can also be used to construct pulsar-based timescales and test Solar System ephemerides \citep{2013CQGra..30v4007H}.

The integrated pulse profile of a pulsar is highly stable, similar to a human fingerprint, but the shape of the individual pulses that make up the integrated profile can vary significantly. The emission variations in individual pulses usually include mode changing \citep[e.g.,][]{1968Natur.220..231D}, nulling \citep[e.g.,][]{1970Natur.228.1297B}, subpulse drifting \citep[e.g.,][]{1968Natur.220..231D}, periodic amplitude modulation \citep[e.g.,][]{2025ApJ...992..105B}, giant pulses \citep[e.g.,][]{2007Ap&SS.308..563K}, and so on.
The individual pulse emission variation phenomena have been widely studied in canonical pulsars \citep[e.g.,][]{2019MNRAS.482.3757B,2020ApJ...889..133B,2023MNRAS.520.4562S,2024ApJ...968..119X},
However, only a few MSPs have shown evidence of mode changing \citep[e.g.,][]{2018ApJ...867L...2M,2020ApJ...902L..13W,2021ApJ...913...67W,2022MNRAS.510.5908M} or sub-pulse drifting \citep{2002A&A...393..733E,2016MNRAS.463.3239L,2023MNRAS.526.2156M}.
MSPs are rarely detected with single pulse variations due to their short rotation periods and weak radiation. Therefore, studying the single pulse variations of MSPs requires telescopes with high sensitivity and high time resolution. FAST will provide a great opportunity to conduct this research \citep{2024ApJ...964....6W,2025ApJ...982..117X}.

PSR J1857+0943 was discovered by the Arecibo radio telescope in 1986. It has a spin period of 5.362 ms and a white-dwarf companion in a nearly circular orbit with a period of approximately 12.3 days \citep{1986Natur.322..714S,2000ApJ...530L..37V}. The previous studies have not characterized its single pulse variation behavior. In this paper, we have analyzed the single pulse modulation and polarization of PSR J1857+0943 using the high-sensitivity FAST Observation. The details of observations and data processing are presented in Section~\ref{sec:Obs}, and the results are presented in Section~\ref{sec:results}. The conclusion and discussion of the study are presented in Section~\ref{sec:disscussion}. 

\section{Observations and Data Reduction}\label{sec:Obs}

The Five-hundred-meter Aperture Spherical Radio Telescope (FAST) is located in Guizhou province in China. FAST is the most sensitive single-aperture radio telescope in the world, and it has an effective illumination aperture of 300 meters during normal operation \citep{Nan+2011,Jiang+2019}. In this paper, we use the data collected by the center beam of the 19-beam receiver with a covering bandwidth range from 1.0$-$1.5 GHz and recorded by a digital backend based on Reconfigurable Open-architecture Computing Hardware$-$version2 (ROACH2) \citep{Li+2018,Jiang+2019,JTH+2020RAA}. The data was recorded in search mode and with the data format as \textsc{PSRFITS} \citep{2004PASA...21..302H}. We chose to record 4 polarizations and 8-bit samples. The sampling interval is 8 $\mu s$, frequency channelization is 1024, and the  observation duration is 1726s. To calibrate the data, our observation contains a stable noise signal injected by a single diode. 

After obtaining the data, we first processed the data using the \textsc{dspsr} software package \citep{2011PASA...28....1V} and eliminated the dispersion delays in the inter-channel using the $-K$ option. Here, the pulsar ephemeris was obtained from the Australia Telescope National Facility(ATNF) pulsar catalogue (PSRCAT V-2.51 \citep{2005AJ....129.1993M} \footnote{http://www.atnf.csiro.au/research/pulsar/psrcat/}). We use 512 phase bins to achieve high time resolution in the pulse profile. 
The folded phase-bin duration of approximately $10.473~\mu{\rm s}$ is not an integer multiple of the raw sampling interval of $8.192~\mu{\rm s}$. Individual phase bins within a single rotation may therefore contain different numbers of raw samples and have different radiometer-noise levels. As the sampling grid shifts relative to rotational phase between successive pulses, the bins receiving an additional sample vary from pulse to pulse. Over the full set of 322,046 rotations, this redistribution is expected to reduce persistent differences in the sampling-related noise level between phase bins.
Secondly, the narrow-band radio frequency interference (RFI) and 50 MHz at each band edge were removed by using the {\sc paz} package provided by the \textsc{PSRCHIVE} software \citep{2004PASA...21..302H}.  After the band-edge removal, the analyzed frequency range was 1.05--1.45 GHz, corresponding to a frequency span of 400 MHz. Finally, the polarization was calibrated by using the \textsc{pac} package of \textsc{PSRCHIVE} software with the noise signal. 
The rotation measure (RM) was determined to be $33.32~\mathrm{rad,m^{-2}}$ using the {\sc rmfit} in {\sc PSRCHIVE}.
The resulting RM is $33.32 \pm 1.52~{\rm rad~m^{-2}}$. The contribution from the Earth's ionosphere was estimated using {\sc ionFR} \citep{2013A&A...552A..58S}, giving $4.01 \pm 0.26~{\rm rad~m^{-2}}$. After correcting for this ionospheric contribution, the rotation measure originating from the interstellar medium is $29.31 \pm 1.54~{\rm rad~m^{-2}}$.
This interstellar RM value was subsequently adopted for all polarization calibration and position angle (PA) corrections applied in our analysis.

\section{RESULTS}\label{sec:results}

\begin{figure}
\centering
\includegraphics[width=3.5in,height=4in,angle=0]{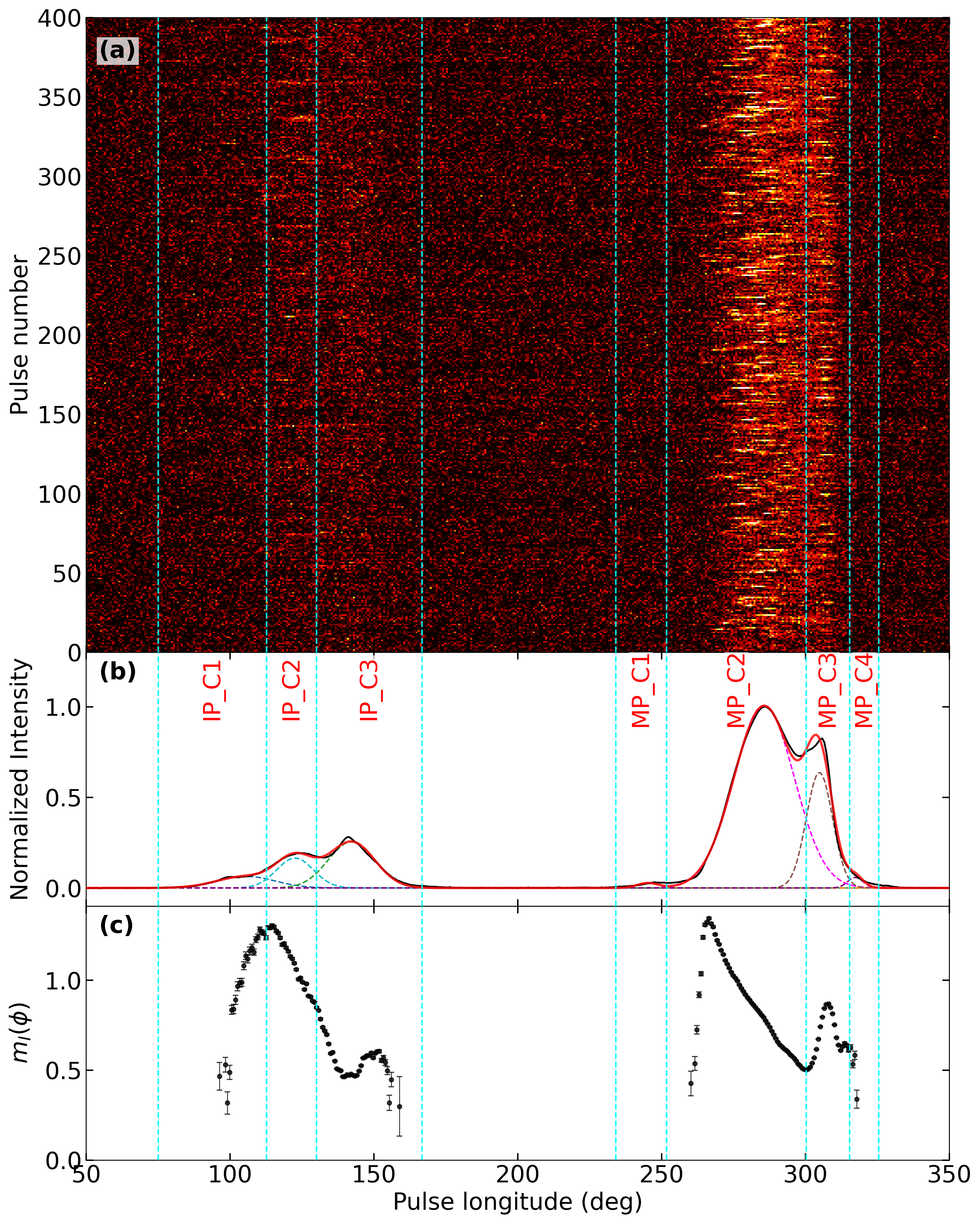}
\caption{Single-pulse sequence and integrated pulse profile for PSR J1857+0943. 
Panel (a) shows the intensity of the first 400 pulses as a function of pulse longitude and pulse number.
Panel (b) displays the integrated profile from all 322,046 pulses, together with the best-fitting Gaussian model and the individual Gaussian components. The main pulse (MP) is resolved into four components, MP\_C1--MP\_C4, while the interpulse (IP) is resolved into three components, IP\_C1--IP\_C3. Panel (c) shows the phase-resolved modulation index computed from the full set of 322,046 pulses, displayed only for on-pulse longitude bins where the average profile is significantly detected. The cyan dashed lines mark the boundaries of the different profile components derived from the Gaussian decomposition.}
\label{fig:J1857+0943_single_pulse_stacks}
\end{figure}

\begin{figure*}
\centering
\includegraphics[width=3.5in,height=3in]{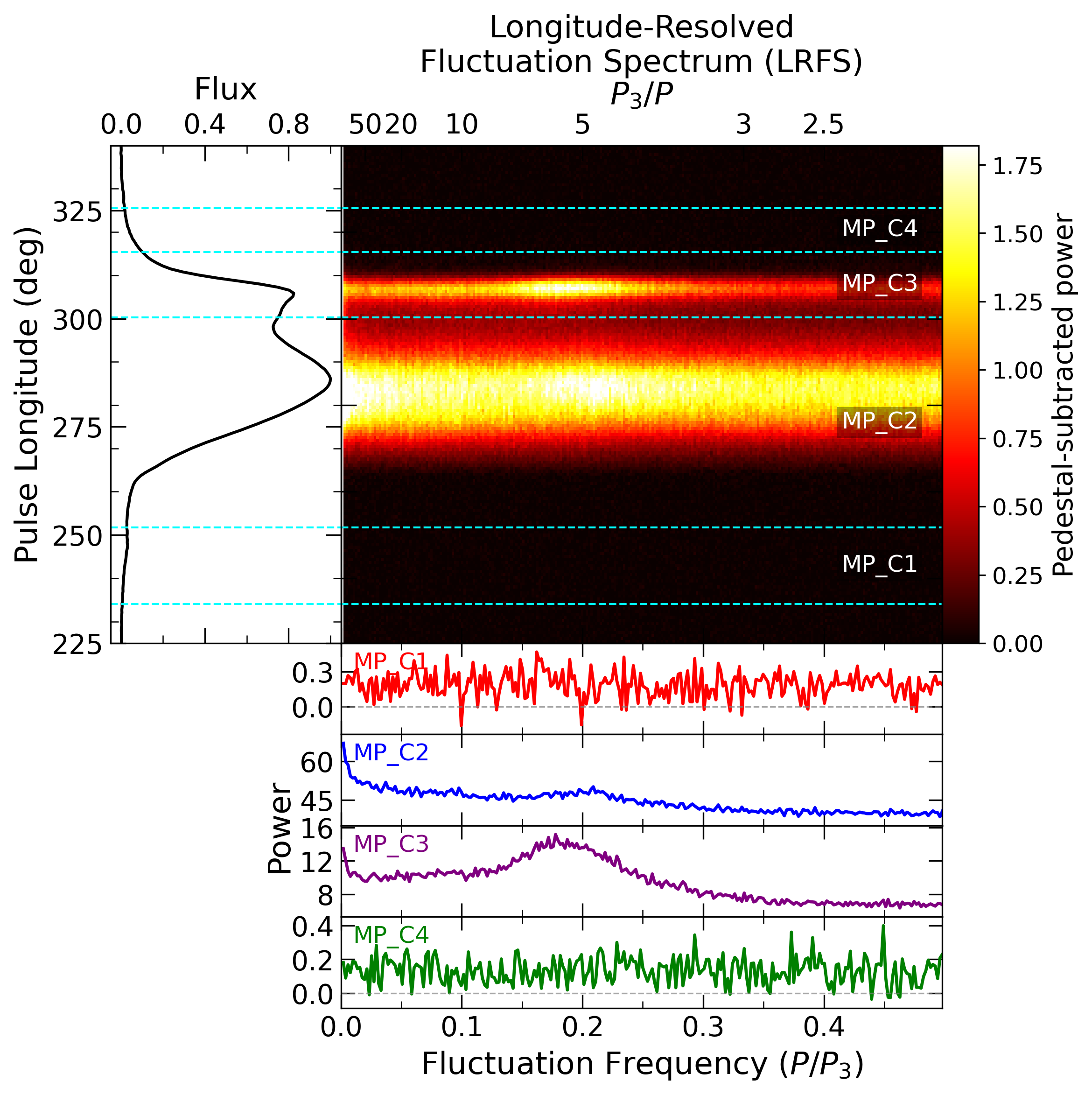}
\includegraphics[width=3.5in,height=3in]{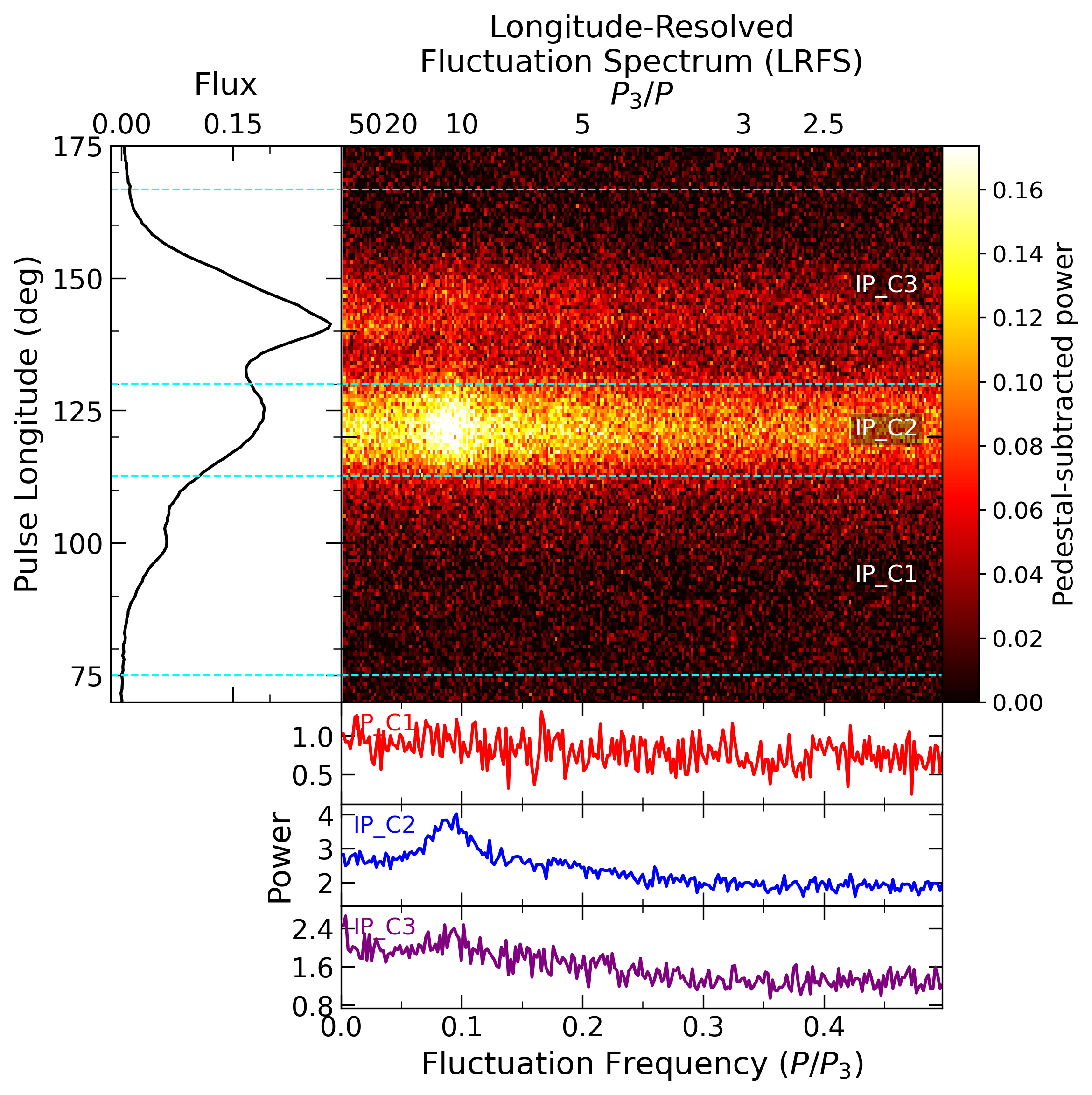}
\caption{Longitude-resolved fluctuation power spectra for the main pulse (left) and interpulse (right) of PSR J1857+0943. The main panel in each plot shows the LRFS, the left panel displays the integrated pulse profile, and the bottom panel shows the total fluctuation power as a function of $f_3=P/P_{3}$. The total power spectrum in the bottom panel is obtained by vertically integrating the LRFS within the longitude ranges enclosed by the cyan dashed lines. For the MP, the red, blue, purple, and green curves represent the total fluctuation-power spectra of MP\_C1, MP\_C2, MP\_C3, and MP\_C4, respectively. For the IP, the red, blue, and purple curves correspond to IP\_C1, IP\_C2, and IP\_C3, respectively.}
\label{fig:LRFS}
\end{figure*}

\subsection{Periodic emission variation}

In total, 322,046 individual pulses from PSR\,1857+0943 were analyzed.  
In order to analyze the individual pulse properties of pulsars, it is customary to reformat the de-dispersed one-dimensional time series into a two-dimensional pulse stack, mapping pulse longitude as a function of pulse number. The upper panel of Figure~\ref{fig:J1857+0943_single_pulse_stacks} presents the pulse stack for the first 400 single pulses, where evident intensity modulation can be seen from pulse to pulse. By vertically integrating the pulse stack over the full dataset, we obtained the integrated pulse profile, as shown in the middle panel of Figure~\ref{fig:J1857+0943_single_pulse_stacks}. The profile clearly reveals a prominent main pulse (MP) and a distinct interpulse (IP). A Gaussian decomposition of the integrated profile identifies four components in the MP (MP\_C1--MP\_C4) and three components in the IP (IP\_C1--IP\_C3).

\subsubsection{The modulation index}

To quantitatively characterize the observed pulse-to-pulse intensity variation, we calculated the longitude-resolved modulation index $m$, defined as 
\begin{equation}
   m= \frac{\sigma_{\text{pulse}}}{\mu_{\text{pulse}}}
\end{equation}
where $\sigma_{\text{pulse}}$ and $\mu_{\text{pulse}}$ are the intrinsic pulse standard deviation and mean pulse intensity at each longitude bin, after correcting for the contribution of the radiometer noise from the off-pulse region. The calculation of $m_I(\phi)$ in this work is follows the standard formulation, with corrections applied for the contribution of radiometer noise estimated from the off-pulse region. In our analysis, the baseline level of each pulse was determined using a single off-pulse window consisting of $N$ phase bins. The statistical effect of this baseline subtraction introduces a small bias in the measured variances, which is explicitly corrected. The final expression for the modulation index is given by
\begin{equation}
m = \frac{\sigma_{\text{pulse}}}{\mu_{\text{pulse}}}
= \frac{ \sqrt{\sigma_{\text{on}}^2 - \frac{N+1}{N-1} \sigma_1^2} }{ \mu_{\text{on}} },
\end{equation}
where $\sigma_{\text{on}}$ and $\mu_{\text{on}}$ denote the standard deviation and mean of the on-pulse region after baseline subtraction, $\sigma_1$ represents the standard deviation of the off-pulse region after baseline subtraction, and $\frac{N+1}{N-1}$ is the correction factor accounting for the statistical coupling introduced by the finite off-pulse window.
A detailed derivation of this correction is presented in  Appendix~\ref{App:Appendux_A}

The off-pulse variance does not include the possible increase
in radiometer noise caused by the pulsar signal itself. We
therefore assessed the importance of pulsar self-noise by
estimating the single-pulse peak flux density relative to the
instrumental system-equivalent flux density, defined as
\begin{equation}
{\rm SEFD}=\frac{T_{\rm sys}}{G},
\end{equation}
where $T_{\rm sys}$ and $G$ are the system temperature and
telescope gain, respectively.

For each pulse, we defined the peak-to-noise ratio as
\begin{equation}
({\rm S/N})_{\rm peak}
=
\frac{A_{\rm peak}}{\sigma_{\rm off}},
\end{equation}
where $\sigma_{\rm off}$ is the rms of the baseline-subtracted off-pulse samples and $A_{\rm peak}$ is the peak amplitude. 

Using the radiometer equation, the dimensionless peak-flux
ratio is
\begin{equation}
\frac{S_{\rm peak}}{{\rm SEFD}}
=
\frac{
\beta({\rm S/N})_{\rm peak}
}{
\sqrt{
n_{\rm pol}\Delta\nu t_{\rm bin}
}
}
=
\frac{
\beta A_{\rm peak}
}{
\sigma_{\rm off}
\sqrt{
n_{\rm pol}\Delta\nu t_{\rm bin}
}
},
\end{equation} 
where $n_{\rm pol}=2$, $\Delta\nu=400$ MHz is the effective bandwidth adopted in this estimate, $t_{\rm bin}=10.473~\mu\mathrm{s}$ is the folded phase-bin duration, and $\beta=1$ corresponds to the ideal-radiometer approximation.
Because $T_{\rm sys}$ and $G$ cancel in this ratio, $S_{\rm peak}/{\rm SEFD}$ can be estimated directly from the uncalibrated single-pulse data.
For the MP, the median, 99th-percentile, and maximum values of $S_{\rm peak}/{\rm SEFD}$ are 0.024, 0.051, and 0.091, respectively. The corresponding values for the IP are 0.007, 0.021, and 0.047. Thus, the overwhelming majority of the pulses remain well within the weak-source regime, and even the brightest measured template amplitudes are only approximately $0.1\,{\rm SEFD}$. Pulsar self-noise is therefore not a dominant contribution to the measured single-pulse variance.

The longitude-resolved modulation index, shown in Figure~\ref{fig:J1857+0943_single_pulse_stacks}(c), indicates that the strongest pulse-to-pulse variability occurs in the leading parts of IP\_C2 and MP\_C2, where it reaches $m=1.303\pm0.010$ at approximately $114.6^\circ$ and $m=1.344\pm0.007$ at approximately $266.5^\circ$, respectively. The $m>1$ region in MP\_C2 coincides with the cluster of bright single pulses visible in Figure~\ref{fig:J1857+0943_single_pulse_stacks}(a). Toward the trailing edge of MP\_C2, the modulation index decreases to $m=0.501\pm0.001$ at approximately $300.2^\circ$, close to the MP\_C2--MP\_C3 boundary, where comparably bright single pulses are largely absent and the emission is dominated by broader, fainter structures. It subsequently rises to a local maximum of $m=0.868\pm0.002$ near the center of MP\_C3 at approximately $308.0^\circ$. The substantial overlap between the adjacent phenomenological components is likely to contribute to the reduced modulation index near the MP\_C2--MP\_C3 boundary through the superposition of emission with different variability properties. Other explanations cannot be ruled out with the present data.

\subsubsection{LRFS}

To investigate the subpulse modulation behavior, we computed the longitude-resolved fluctuation spectrum (LRFS) following the standard fluctuation analysis technique \citep{1970Natur.228.1297B,2002A&A...393..733E}. 
The pulse sequence was divided into 628 non-overlapping blocks of 512 pulses, with the remaining 510 pulses excluded from the spectral calculation.
For each block, only the on-pulse longitude bins were selected. To eliminate the DC component of the fluctuation spectrum, the mean intensity at each pulse longitude was subtracted. A Fast Fourier Transform (FFT) was then performed along the pulse number axis for every longitude bin, converting temporal intensity variations into modulation power as a function of fluctuation frequency. The final LRFS was obtained by averaging the power spectra over all data blocks. Figure~\ref{fig:LRFS} shows the resulting LRFS for the MP and IP of PSR J1857+0943.

As shown in Figure~\ref{fig:LRFS}, the emission components exhibit distinct periodic modulation properties. No distinct localized periodic feature is detected in MP\_C1 or MP\_C4. MP\_C2 shows a relatively weak feature, whereas MP\_C3 exhibits a prominent feature at approximately $f_3=P/P_3\simeq0.19$ cpp, corresponding to $P_3\simeq5.3P$. For the IP, no distinct periodic feature is detected in IP\_C1. IP\_C2 exhibits a prominent feature and IP\_C3 shows a weaker feature at approximately $f_3\simeq0.09$ cpp, corresponding to $P_3\simeq11.1P$. The absence of significant peaks in MP\_C1, MP\_C4, and IP\_C1 suggests a lack of stable, periodic subpulse modulation in these regions.

\begin{figure*}
    \centering
    \includegraphics[width=2.2in]{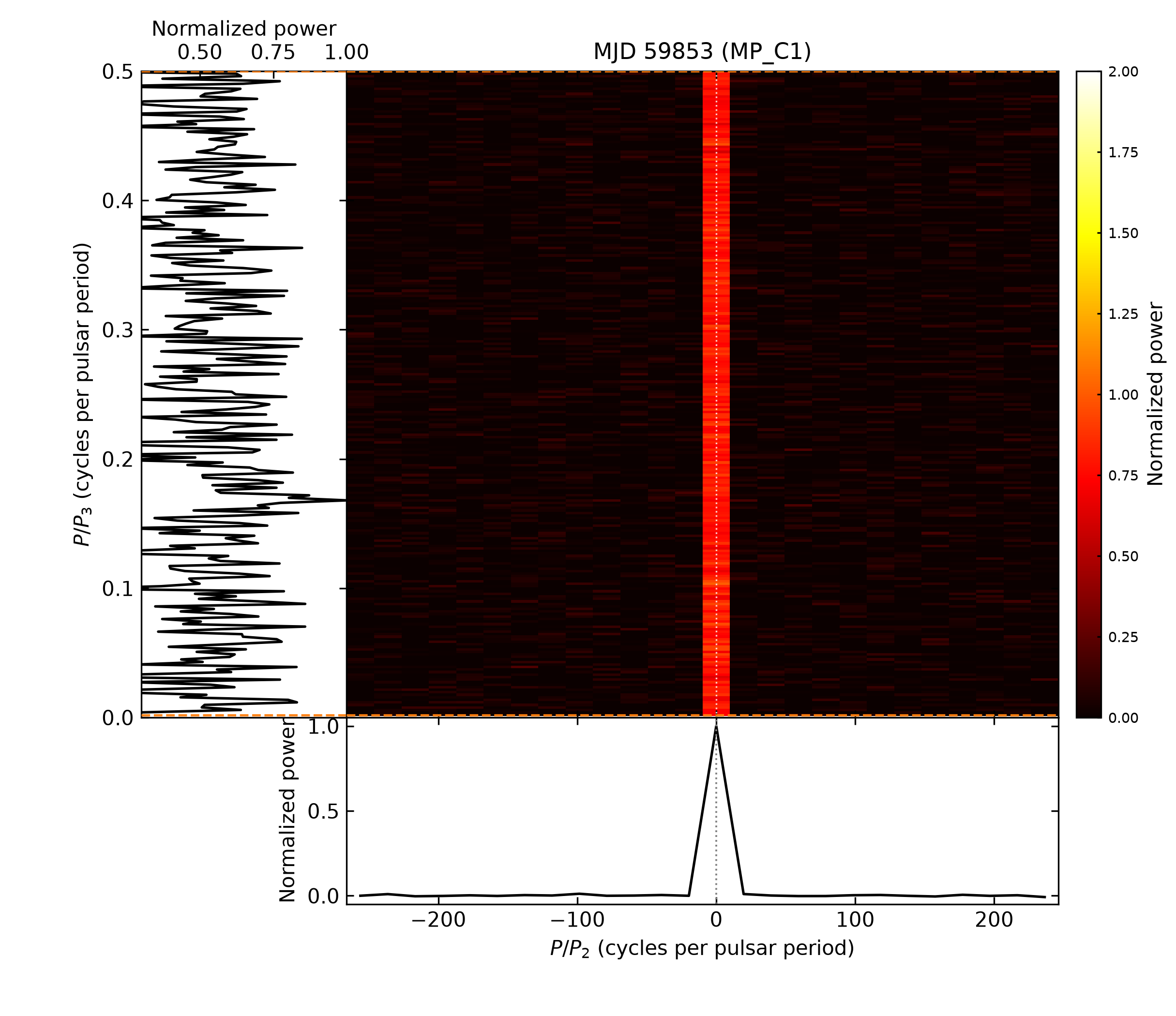}
    \includegraphics[width=2.2in]{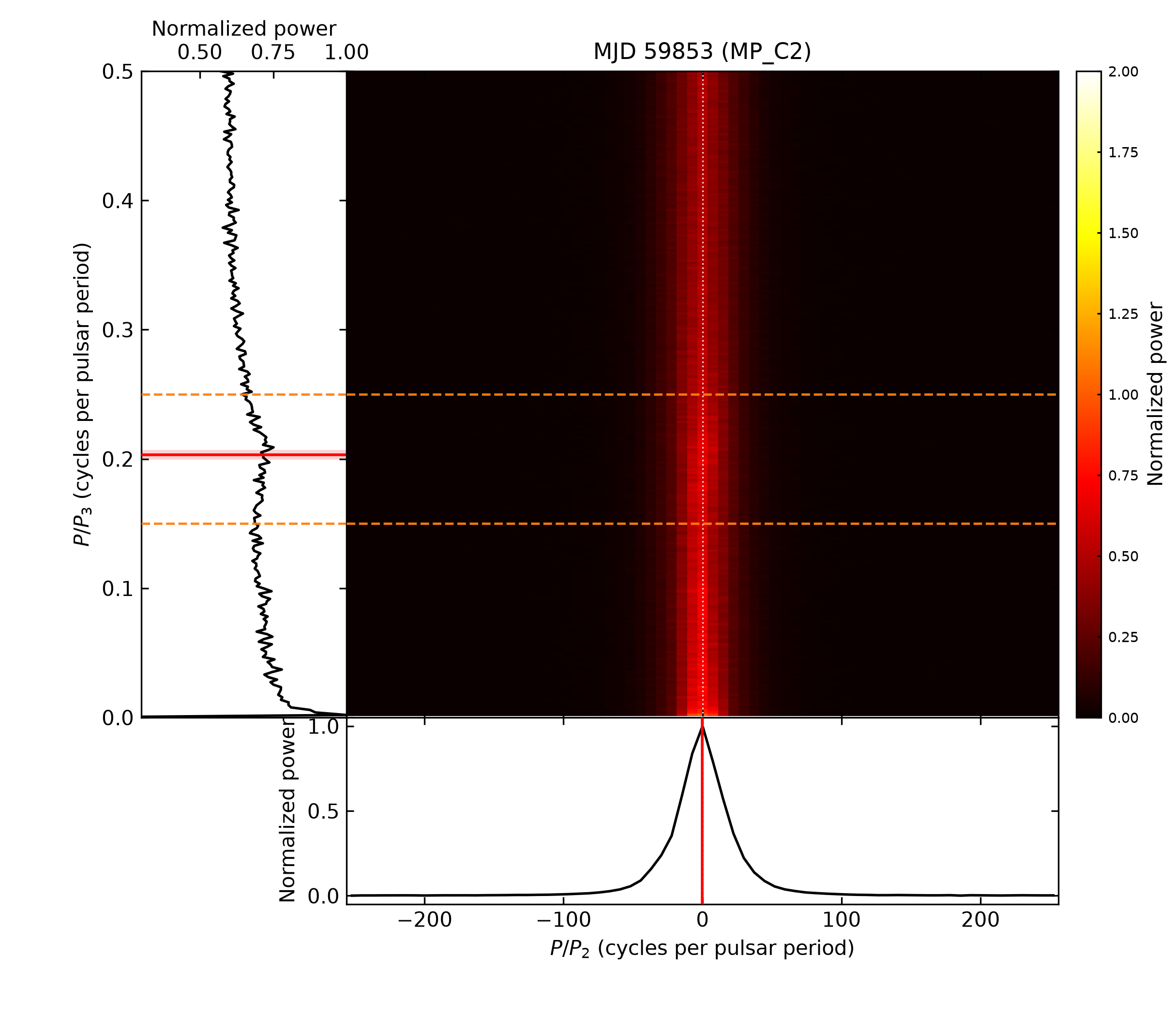}
    \includegraphics[width=2.2in]{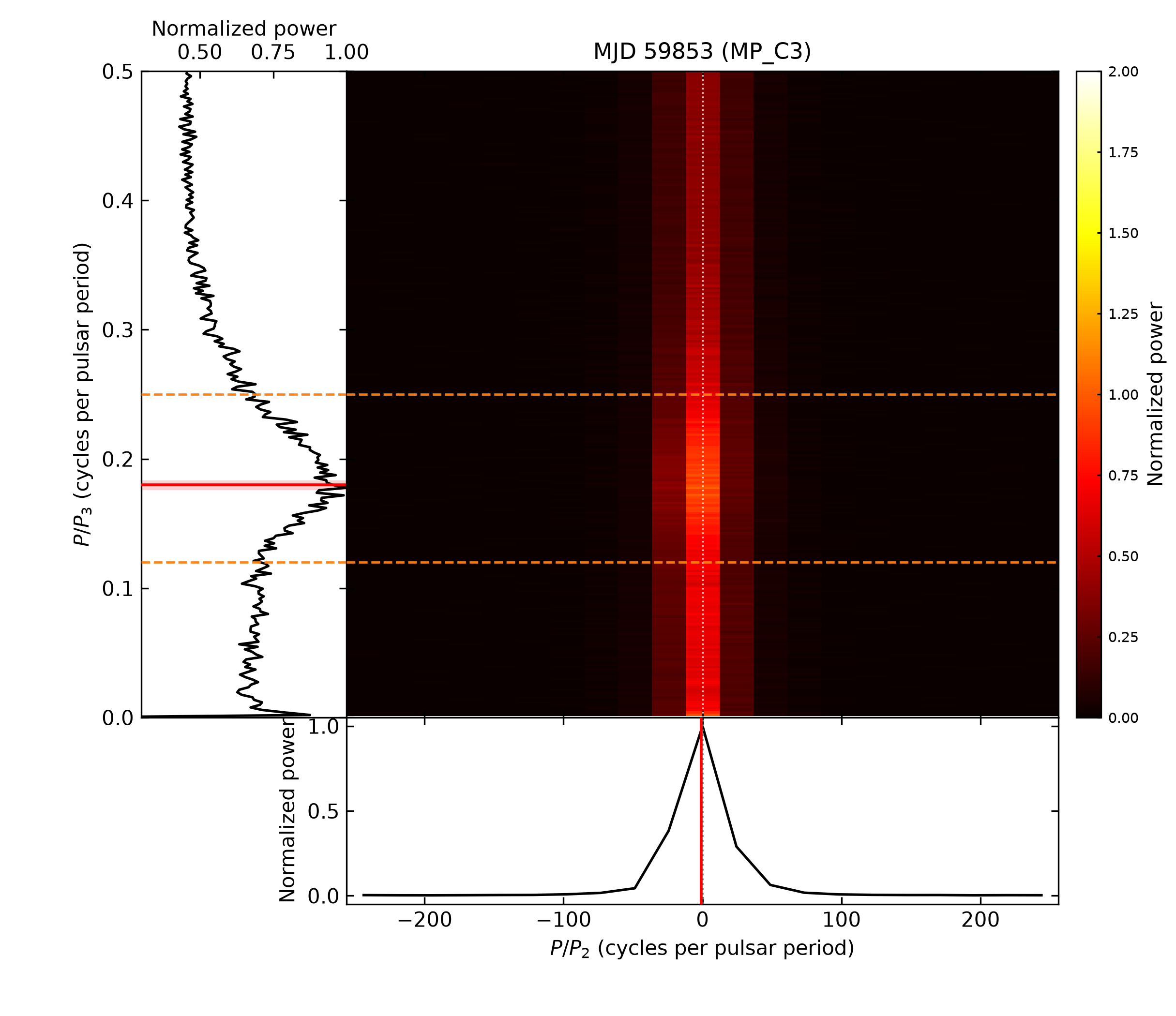}
    \includegraphics[width=2.2in]{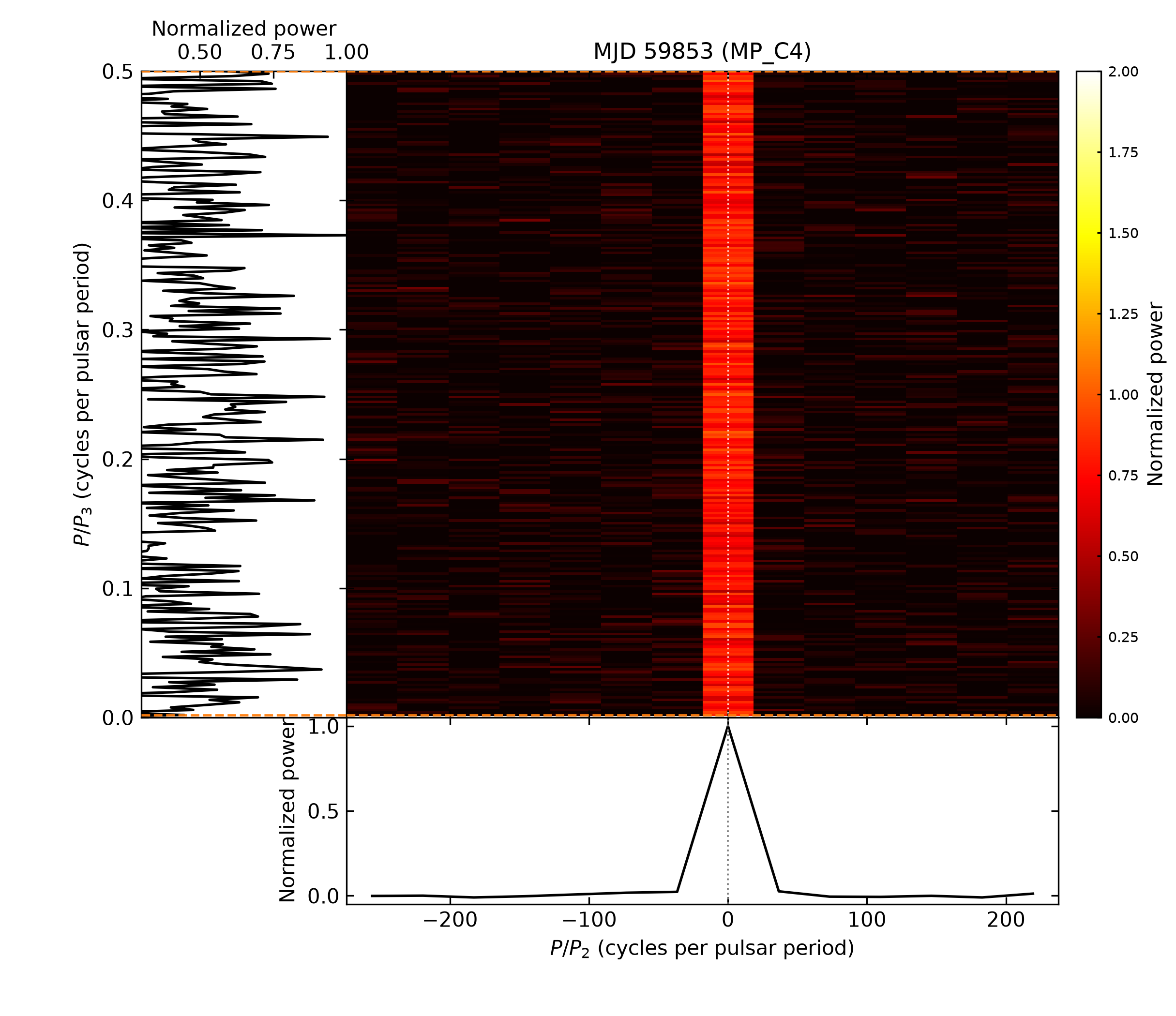}
    \includegraphics[width=2.2in]{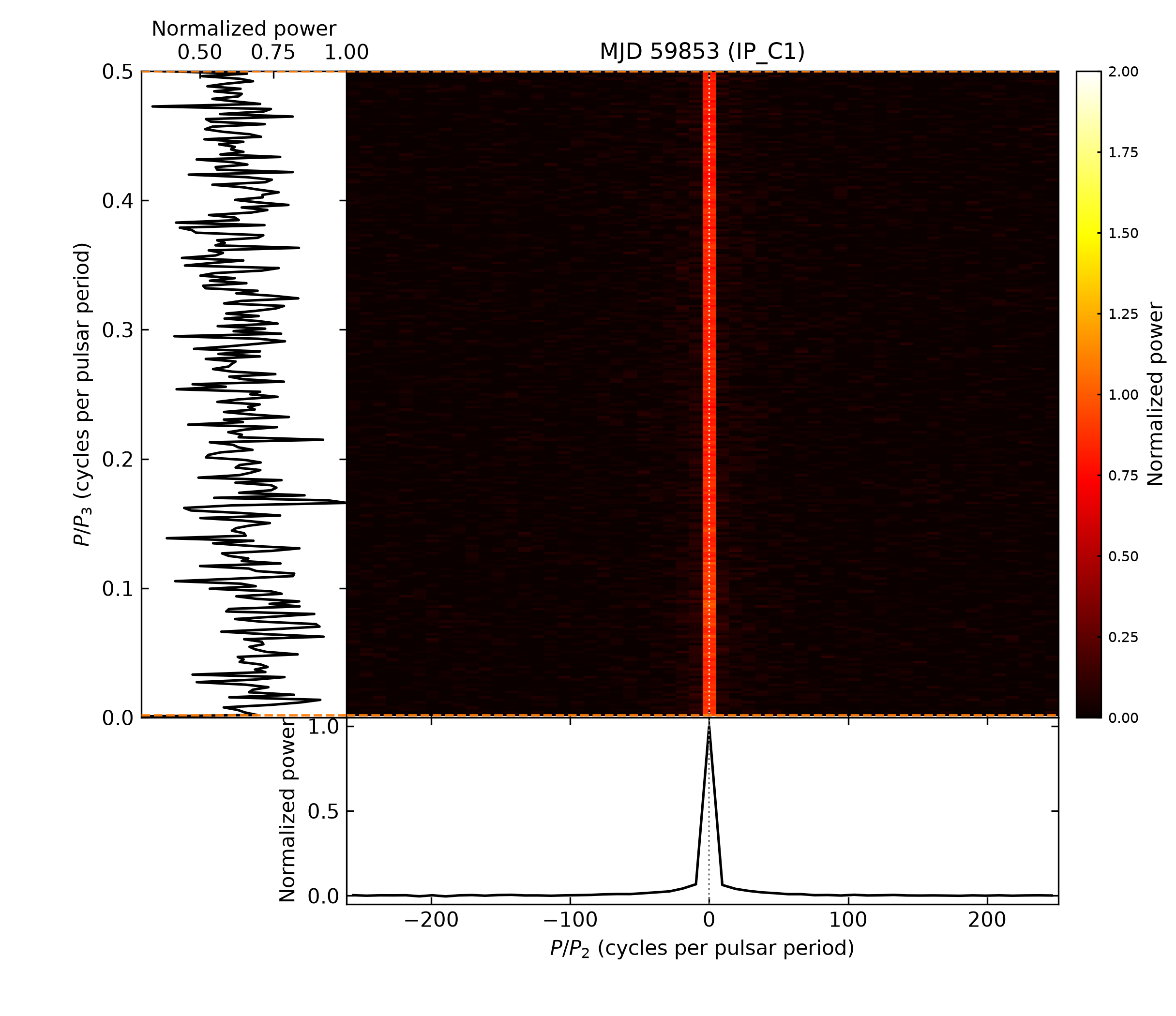}
    \includegraphics[width=2.2in]{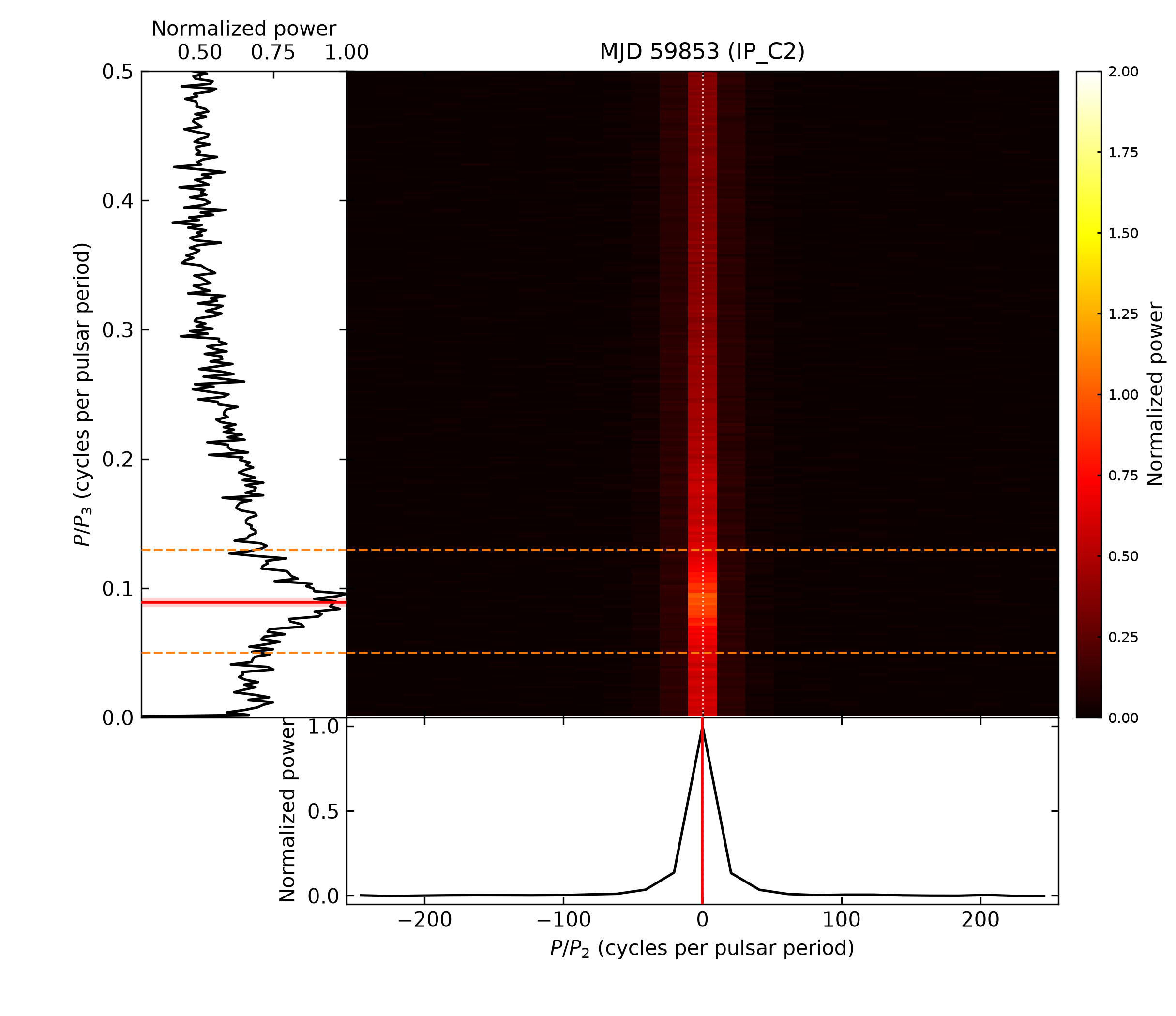}
    \includegraphics[width=2.2in]{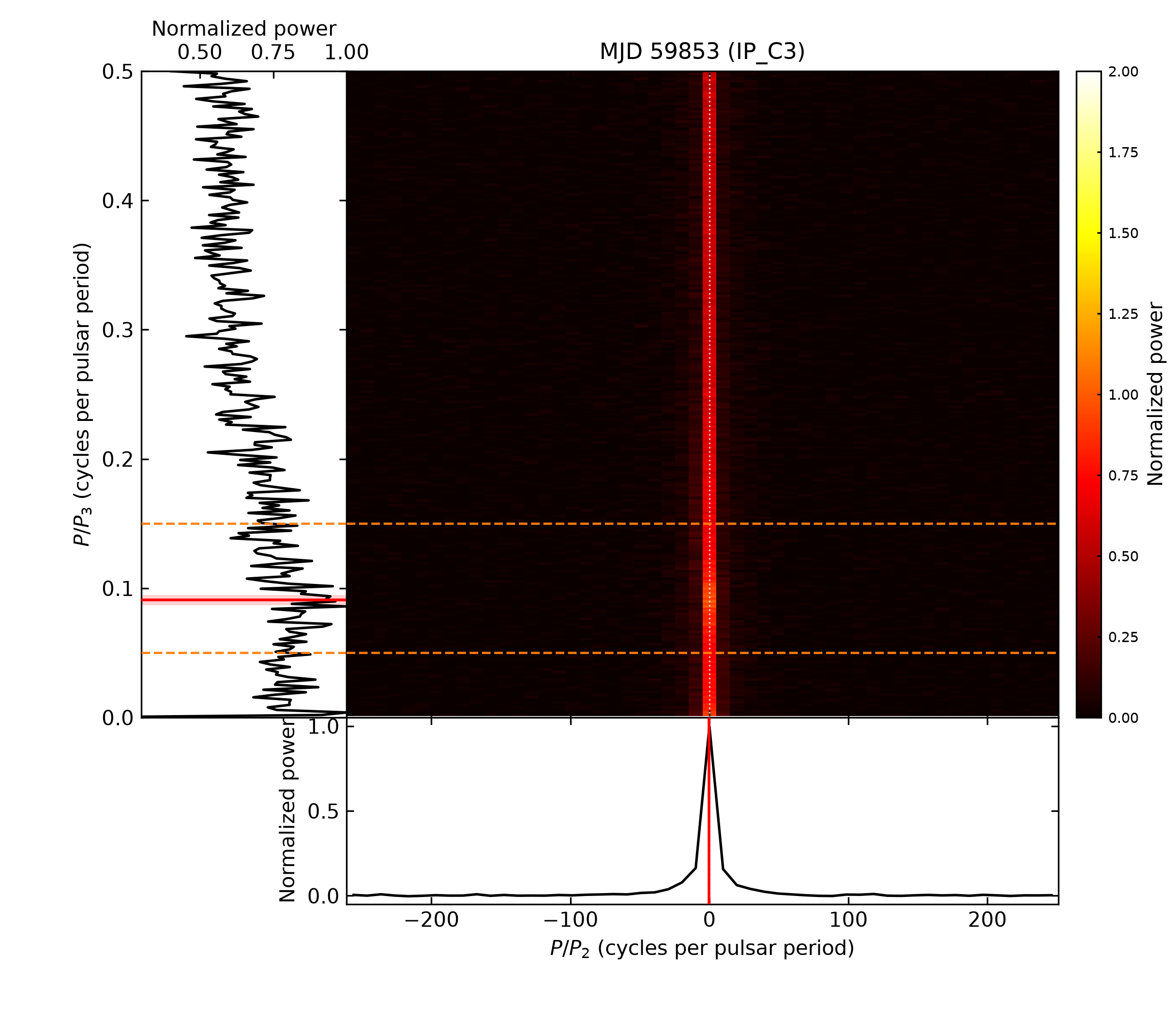}
\caption{Component-resolved two-dimensional fluctuation spectra (2DFS) of PSR J1857+0943 at MJD 59853. In each panel, the central color map shows the 2DFS.  The black curve in the left panel is the $P/P_3$ power spectrum obtained by integrating over the displayed $P/P_2$ range, while the black curve in the bottom panel is the $P/P_2$ power spectrum integrated over the $P/P_3$ interval bounded by the orange dashed lines. The white dotted vertical line in the central 2DFS and the gray dotted vertical line in the lower projection mark $P/P_2=0$. For IP\_C2, IP\_C3, MP\_C2, and MP\_C3, the red solid lines in the marginal spectra mark the power-weighted centroid frequencies, and the light-red shaded regions show the uncertainty envelopes obtained by varying the selected 2DFS rectangle and including the corresponding 1$\sigma$ white-noise centroid uncertainty.}
\label{fig:J1857+0943_2dfs}
\end{figure*}

\subsubsection{2DFS}

Although the longitude-resolved fluctuation spectrum reveals periodic pulse-to-pulse modulation, it cannot distinguish stationary amplitude modulation from systematic subpulse-phase drift. We therefore calculated the two-dimensional fluctuation spectra (2DFS) for the seven profile components, MP\_C1$-$MP\_C4 and IP\_C1$-$IP\_C3, following \citet{2002A&A...393..733E}.The vertical and horizontal fluctuation frequencies are denoted by $f_3=P/P_3$ and $f_2=P/P_2$, respectively, both in cpp.
 
The pulse sequence was divided into 628 non-overlapping blocks of 512 pulses, and the remaining 510 pulses were excluded. For each block, a DFT was first applied along pulse number and then along pulse longitude. The resulting two-dimensional power spectra were then averaged over all blocks.
An off-pulse 2DFS with the same longitude-window width, calculated using the identical procedure, was subtracted from each on-pulse spectrum. The $f_3=0$ row was excluded from the subsequent analysis.
The modulation features were characterized using power-weighted centroids within selected rectangular regions \citep{2006A&A...445..243W,2016A&A...590A.109W}. The quoted uncertainty ranges account for variations in the rectangle boundaries and the corresponding $1\sigma$ white-noise uncertainty in the centroid. They are therefore sensitivity envelopes rather than formal confidence intervals.

As shown in Figure~\ref{fig:J1857+0943_2dfs}, 
no sufficiently isolated $f_3$ feature was identified in IP\_C1, MP\_C1, or MP\_C4 for quantitative measurement. In contrast, localized modulation features are detected in MP\_C2, MP\_C3, IP\_C2, and IP\_C3. 
The MP\_C2 and MP\_C3 features have centroid frequencies of $f_3=0.203$ and $0.179$ cpp, respectively, corresponding to $P_3=4.922^{+0.095}_{-0.091}P$ for MP\_C2 and $P_3=5.560^{+0.119}_{-0.111}P$ for MP\_C3. The IP\_C2 and IP\_C3 features have similar centroid frequencies of $f_3=0.089$ and $0.091$ cpp, corresponding to $P_3=11.24^{+0.48}_{-0.44}P$ and $P_3=11.00^{+0.50}_{-0.44}P$, respectively. Thus, the characteristic modulation period of the two IP components is approximately twice that of MP\_C2 and MP\_C3.

For all four measured features, the maximum-power pixel lies at $f_2=0$. Although their power-weighted centroids show small horizontal offsets, these offsets are less than 0.04 of the corresponding horizontal Fourier-bin width and are therefore not spectrally resolved. Consequently, no finite $P_2$ or preferred drift direction can be reliably determined. The 2DFS results thus demonstrate organized periodic intensity modulation in IP\_C2, IP\_C3, MP\_C2, and MP\_C3, but do not provide evidence for systematic subpulse drifting.

\begin{figure*}
    \centering
    \includegraphics[width=3.2in]{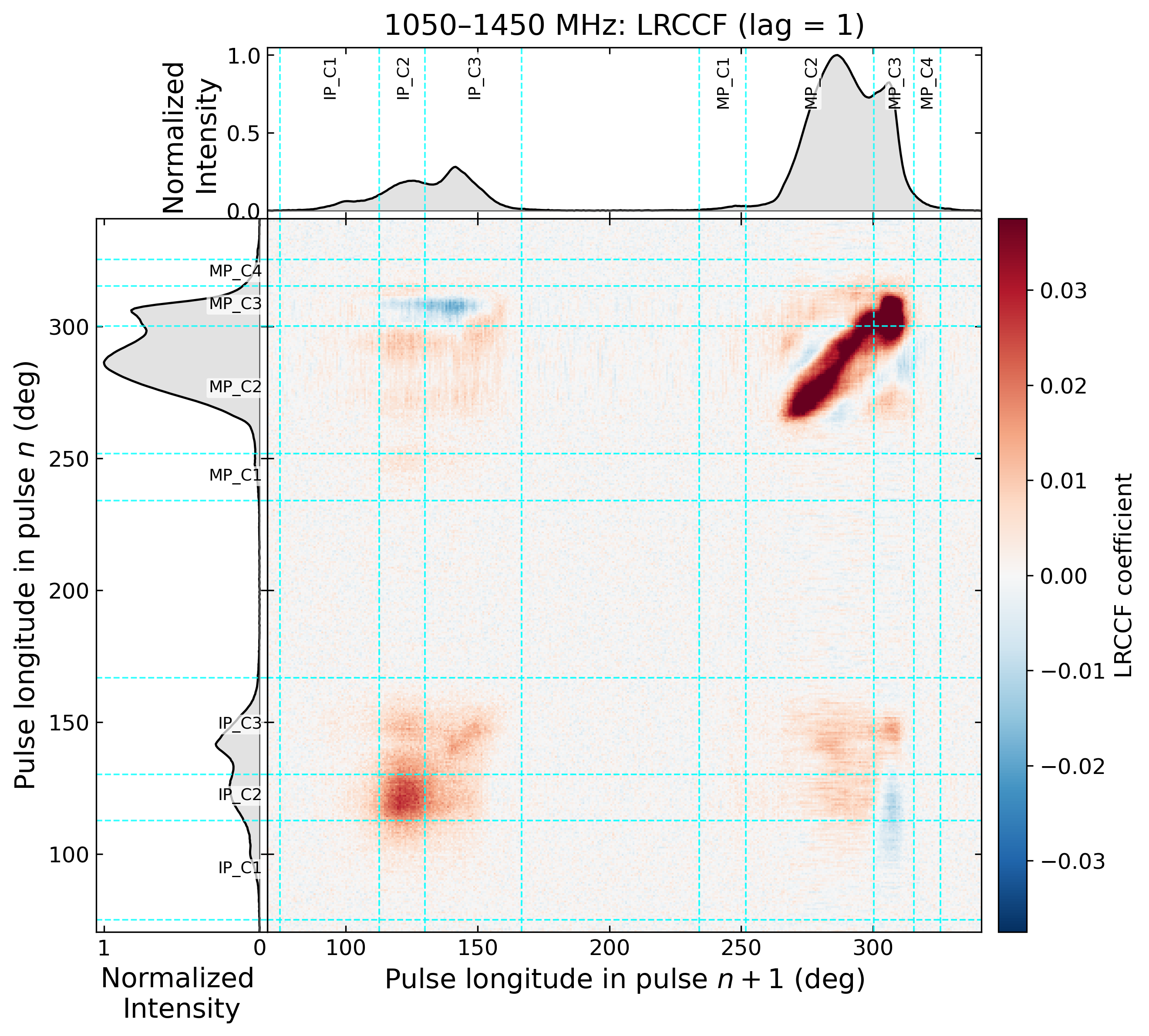}
    \includegraphics[width=3.2in]{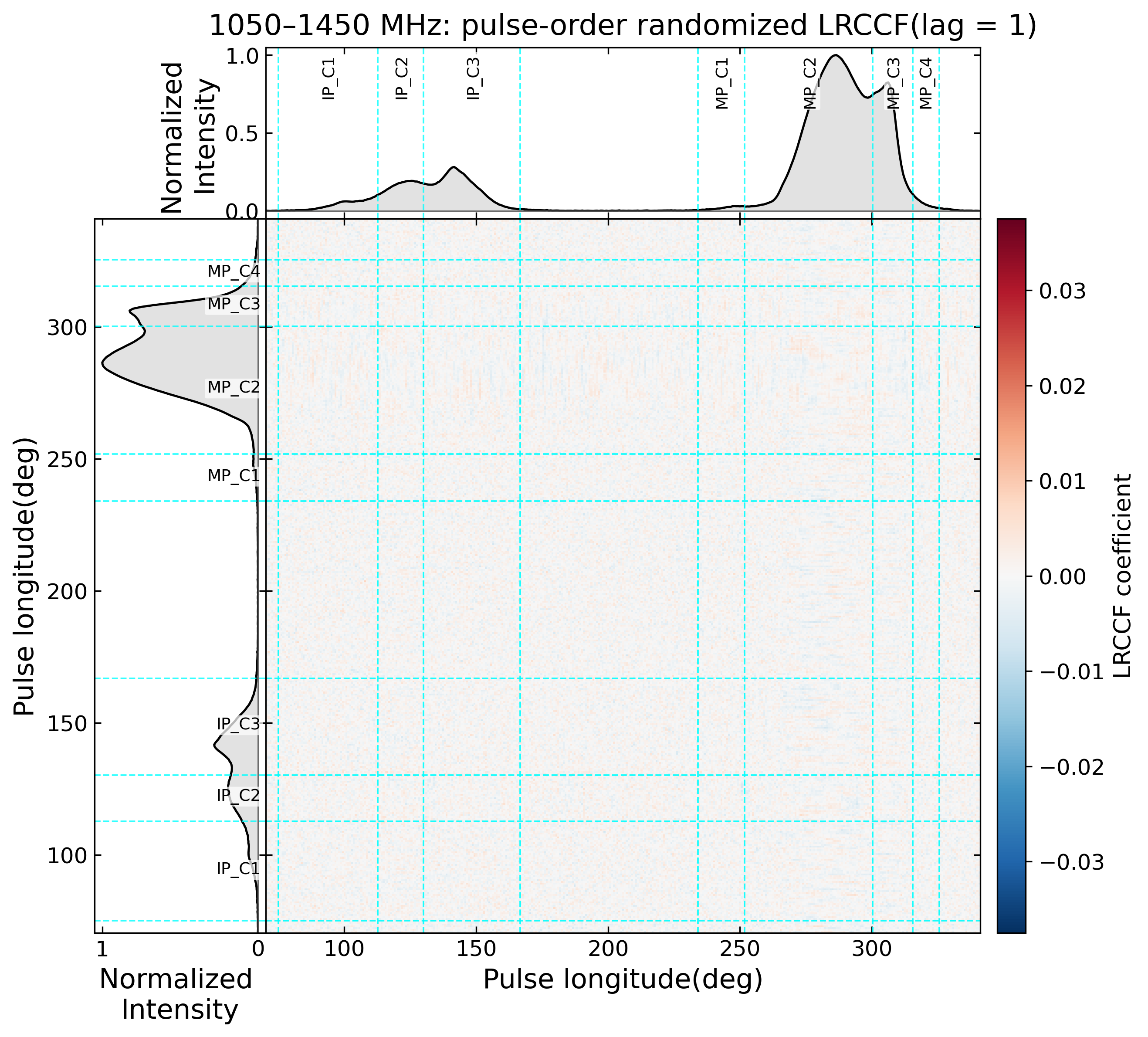}
    \caption{Comparison between the observed and pulse-order-randomized LRCCFs of PSR~J1857+0943 over 1050--1450 MHz at a pulse-number lag of $k=1$ over the entire 322,046 single pulses. The left panel shows the LRCCF calculated from the original pulse sequence, while the right panel shows one representative realization after randomly permuting the pulse order. The cyan dashed lines mark the boundaries of the different profile components.}
    \label{fig:lrccf}
\end{figure*}

\begin{figure*}
    \centering
    \includegraphics[width=7in]{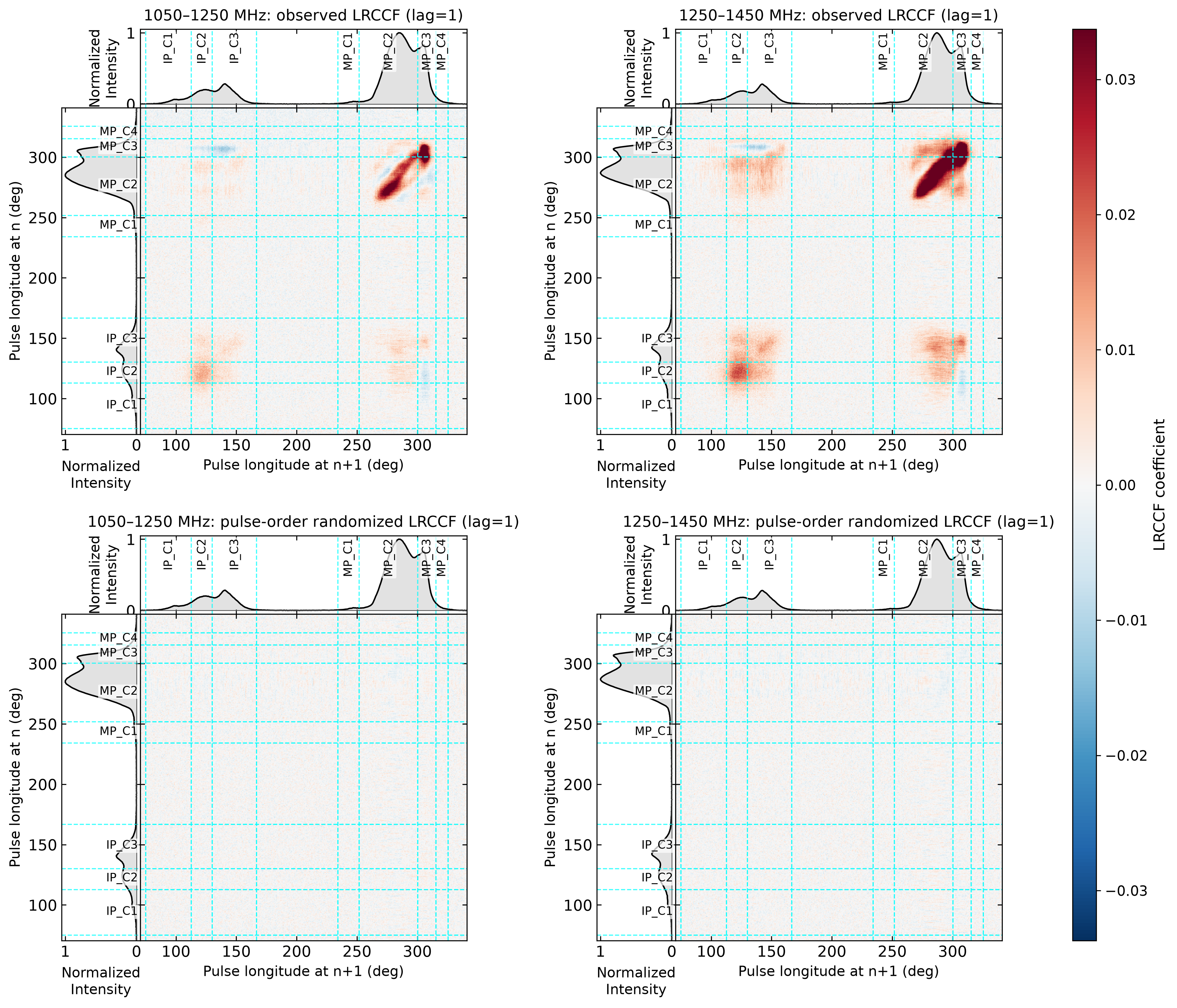}
    \caption{Frequency-resolved longitude-resolved cross-correlation functions (LRCCFs) of PSR~J1857+0943 at a pulse-number lag of $k=1$ over the entire 322,046 single pulses. The upper-left and upper-right panels show the LRCCF maps calculated from the observed pulse sequences in the 1050--1250 MHz and 1250--1450 MHz sub-bands, respectively. The lower-left and lower-right panels show representative pulse-order-randomized realizations for the corresponding sub-bands. The top and left marginal panels show the integrated pulse profile calculated separately for the corresponding sub-band, with each sub-band profile normalized to its own peak intensity. The cyan dashed lines mark the component boundaries derived from the Gaussian decomposition of the full-band (1050--1450 MHz) integrated profile and applied consistently to all panels. The common color scale represents the Pearson correlation coefficient.}
    \label{fig:lrccf_two_band}
\end{figure*}

\subsubsection{LRCCF}

As an additional diagnostic of pulse-to-pulse correlations beyond the LRFS and 2DFS analyses, we computed the longitude-resolved cross-correlation function (LRCCF).
The LRCCF measures the normalized correlation between intensity fluctuations at two pulse longitudes as a function of pulse-number lag. 
For two pulse-longitude bins $\phi_i$ and $\phi_j$, the
longitude-resolved cross-correlation function (LRCCF) at a
pulse-number lag $k$ was calculated as
\begin{equation}
C(\phi_i,\phi_j,k)
=
\frac{
\displaystyle
\sum_{n\in\mathcal{S}_k}
\left[I_n(\phi_i)-\bar{I}_{i,k}^{(0)}\right]
\left[I_{n+k}(\phi_j)-\bar{I}_{j,k}^{(+)}\right]
}{
\displaystyle
\sqrt{
\sum_{n\in\mathcal{S}_k}
\left[I_n(\phi_i)-\bar{I}_{i,k}^{(0)}\right]^2
\sum_{n\in\mathcal{S}_k}
\left[I_{n+k}(\phi_j)-\bar{I}_{j,k}^{(+)}\right]^2
}
},
\label{eq:lrccf}
\end{equation}
where $I_n(\phi)$ is the baseline-subtracted intensity at
pulse longitude $\phi$ in pulse $n$. The set $\mathcal{S}_k$
contains the pulse indices for which both pulse $n$ and pulse
$n+k$ are valid, and $N_k=|\mathcal{S}_k|$ is the corresponding
number of valid pulse pairs. The two means are calculated
separately over the overlapping sequences:
\begin{equation}
\bar{I}_{i,k}^{(0)}
=
\frac{1}{N_k}
\sum_{n\in\mathcal{S}_k} I_n(\phi_i),
\qquad
\bar{I}_{j,k}^{(+)}
=
\frac{1}{N_k}
\sum_{n\in\mathcal{S}_k} I_{n+k}(\phi_j).
\end{equation}
This normalization makes $C(\phi_i,\phi_j,k)$ a Pearson
correlation coefficient bounded between $-1$ and $+1$.
For positive $k$, the intensity at $\phi_i$ in pulse $n$ is
correlated with that at $\phi_j$ in pulse $n+k$.

Inclined correlation ridges in an LRCCF map can indicate systematic longitude evolution of the modulation pattern, whereas nearly vertical structures are expected when the fluctuations are dominated by amplitude modulation with little measurable phase drift.
The left panel of Figure~\ref{fig:lrccf} presents the LRCCF map calculated over the full 1050--1450 MHz band at a lag of one pulse period.
An inclined positive-correlation ridge is visible in the MP\_C2 region. However, because the corresponding 2DFS feature does not show a spectrally resolved offset from $f_2=0$, we do not interpret this ridge as robust evidence for systematic subpulse drifting. In contrast, MP\_C3 shows nearly vertical correlation patches without a clear slope, consistent with variability dominated by amplitude modulation.

The full-band LRCCF also contains both positive and negative correlations between the MP and IP longitude ranges. Positive correlations involve parts of MP\_C1 and MP\_C2, whereas prominent negative regions occur between MP\_C3 and both IP\_C1 and IP\_C2, with the stronger anti-correlation associated with IP\_C2. Thus, the MP--IP correlation structure is not restricted to a single pair of longitude ranges. The localized MP\_C3--IP\_C1 and MP\_C3--IP\_C2 anti-correlations indicate that enhanced emission in MP\_C3 tends, on average, to be associated with reduced interpulse emission in the following pulse.

No clear correlation feature involving MP\_C4 is detected in the full-band LRCCF map. Because MP\_C4 is substantially weaker than the other main-pulse components, its pulse-to-pulse fluctuations are more strongly affected by radiometer noise, which can attenuate the measured correlation coefficient. In addition, overlapping contributions from the trailing emission components may partially cancel in the net longitude-resolved correlation. The absence of a detectable MP\_C4 correlation therefore does not necessarily imply that MP\_C4 is intrinsically independent of the other emission components.

To test whether the observed LRCCF structures depend on the temporal ordering of the pulses, we performed a pulse-order randomization test. For each realization, the pulse numbers were randomly permuted using the same permutation at all longitude bins. This procedure preserves the complete longitude structure and intensity distribution of every individual pulse while destroying correlations associated with the ordering of successive pulses. We generated 1000 randomized pulse sequences and recalculated the $k=1$ LRCCF for each realization.
A representative pulse-order-randomized LRCCF realization is shown in the right panel of Figure~\ref{fig:lrccf}. It contains no coherent correlation ridges or MP--IP correlation patches and is consistent with fluctuations around zero. In particular, the MP\_C3--IP\_C1 and MP\_C3--IP\_C2 anti-correlations present in the original pulse sequence are absent after pulse-order randomization. The randomized realizations therefore provide a control showing that the observed structures are not reproduced when the original temporal ordering of the pulses is destroyed.
The pulse-order randomization test does not, by itself, rule out every possible instrumental or propagation-related systematic effect. Nevertheless, it shows that the observed LRCCF structures are not reproduced by random pairing of otherwise unchanged individual pulses.

Because residual dispersion, digitization artifacts, and aliasing may introduce frequency-dependent correlation patterns, we further examined whether the large-scale full-band LRCCF morphology is reproduced across the observing band and whether the localized correlation features vary with frequency. We therefore repeated both the LRCCF calculation and the pulse-order randomization analysis separately in the 1050--1250 MHz and 1250--1450 MHz sub-bands (the upper two panels in Figure~\ref{fig:lrccf_two_band}). The two observed sub-band maps retain broadly similar large-scale structures, with a pixel-by-pixel Pearson correlation coefficient of $r=0.760$ over the analyzed longitude range. Nevertheless, several localized correlation features show appreciable frequency evolution.

In both sub-bands, the localized anti-correlations between MP\_C3 in pulse $n$ and IP\_C1 and IP\_C2 in pulse $n+1$ retain the same negative sign, although their strengths and detailed longitude-resolved morphologies vary with frequency. These
MP\_C3--IP anti-correlations are therefore more stable in sign across the two sub-bands. In contrast, the correlation between MP\_C3 in pulse $n$ and a restricted longitude range within MP\_C2 in pulse $n+1$ changes sign across the observing band. It is negative in the 1050--1250 MHz sub-band but positive in the 1250--1450 MHz sub-band.

Representative pulse-order-randomized LRCCF maps for the two sub-bands show no comparable coherent structures in the two lower panels of Figure~\ref{fig:lrccf_two_band}. The coherent features in the observed maps are therefore not reproduced when the original pairing between successive pulses is destroyed.


\subsection{The averaged polarization profile}

\begin{figure}
\centering
\includegraphics[width=3.2in,height=4in,angle=0]{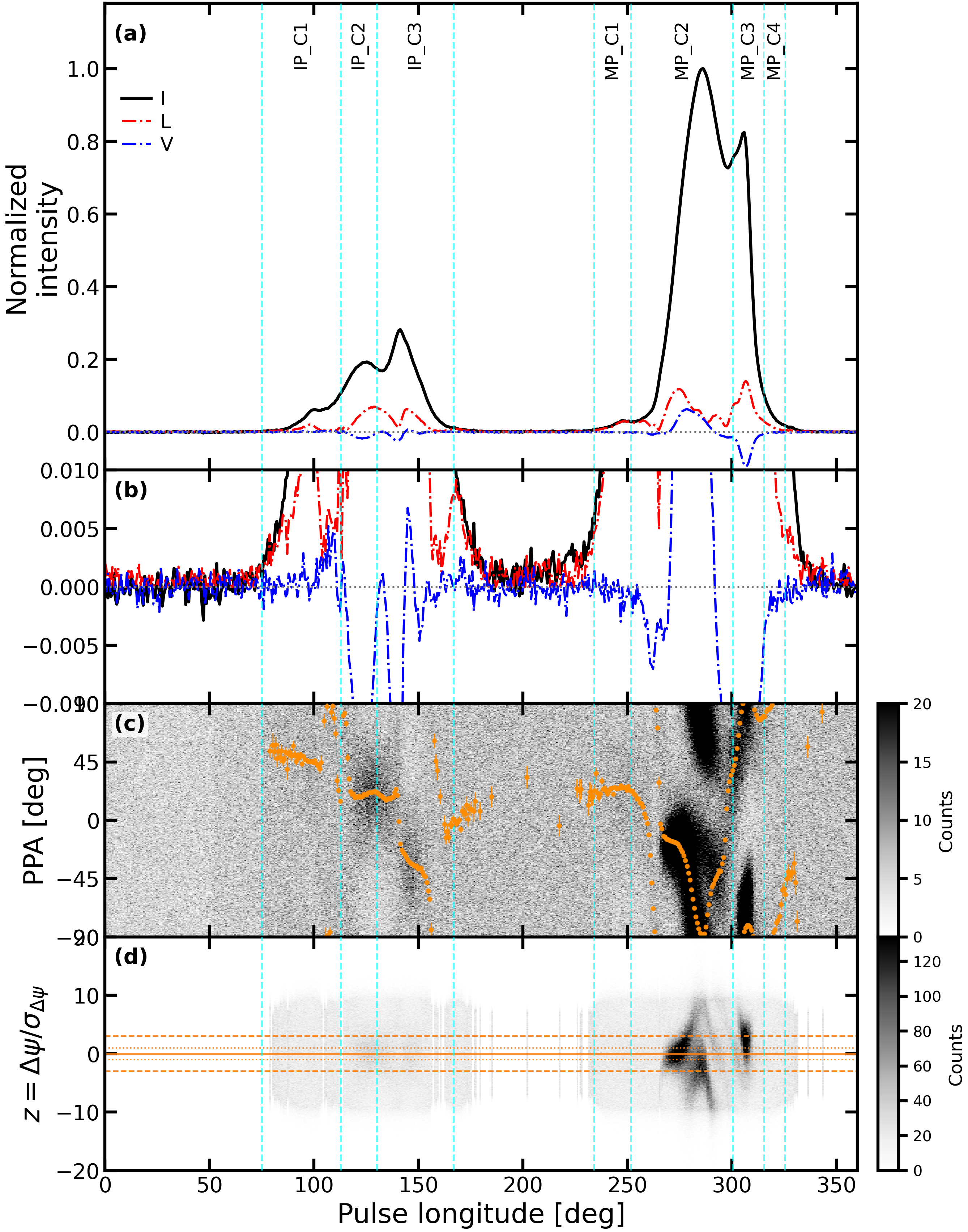}
\caption{Polarization and single-pulse PPA properties of PSR J1857+0943. From top to bottom: (a)the polarized profile, where the black solid line, the red dashed line, and the blue dashed line represent the total intensity (I), linear polarization (L), and circular polarization (V), respectively. (b) Enlarged view of the same three profiles, using the same colors and line styles. (c) Single-pulse PPA distribution for measurements satisfying ($L_{\rm debias}/\sigma_L\geq3$); darker grayscale indicates a larger number of measurements per bin, while the orange points and error bars show the integrated PPA and its uncertainty. (d) Distribution of the uncertainty-normalized PPA residuals, $z=\Delta\psi/\sigma_{\Delta\psi}$. Darker grayscale indicates more measurements per bin, and the orange solid, dotted, and dashed horizontal lines mark $z=0$, $z=\pm1$, and $z=\pm3$, respectively. Gray dotted horizontal lines in panels (a) and (b) mark zero intensity. Cyan dashed vertical lines in all panels mark the component boundaries; the component names are labeled vertically in panel (a). The two color bars give the counts per bin in panels (c) and (d).}
\label{fig:profile_PA}
\end{figure}

The mean pulse profile and polarization properties of a pulsar are important for understanding the pulse emission mechanism, the beaming of pulsar radiation, and the emission geometry. 
The integrated polarization profile and the single-pulse polarization position-angle (PPA) properties of PSR J1857+0943 at 1.25 GHz are shown in Figure~\ref{fig:profile_PA}.
We detect only weak bridge emission in our integrated pulse profile (around a pulse longitude of 180 to 250 degrees), and its significance is lower than the clearer structure reported in the literature \citep{2025A&A...695A.173X}, likely due to the relatively low signal-to-noise ratio of our observation.
To calculate the fractional linear polarization ($\langle L\rangle/I$), the fractional net circular polarization ($\langle V\rangle/I$), and the fractional absolute circular polarization ($\langle |V|\rangle/I$) of the profile, all polarization parameters were derived from the average polarization profile. The on-pulse window was defined as the portion of the pulse profile where the total intensity exceeds three times the baseline rms noise. The measured polarization fractions are $\langle L\rangle/I=14.3\%$, $\langle V\rangle/I=-0.6\%$, and $\langle |V|\rangle/I=4.7\%$, which are in general consistent with the values of $14.2\%$, $-0.1\%$, and $4.3\%$ reported by \citet{2025A&A...695A.173X}.

For the single-pulse analysis, the off-pulse means of Stokes $Q$ and $U$ were subtracted separately from each pulse. The observed and debiased linear-polarization intensities were calculated as $L_{\rm obs}=(Q^2+U^2)^{1/2}$ and $L_{\rm debias}=[\max(L_{\rm obs}^2-\sigma_L^2,0)]^{1/2}$, respectively, where $\sigma_L=[(\sigma_Q^2+\sigma_U^2)/2]^{1/2}$ \citep{2001ApJ...553..341E}. A PPA measurement was retained only when $L_{\rm debias}/\sigma_L\geq3$. The PPA was calculated as $\psi=\frac{1}{2}\operatorname{atan2}(U,Q)$. In the high-linear-polarization-S/N limit, its uncertainty was approximated by $\sigma_\psi=\frac{1}{2}{\sigma_L}/{L_{\rm debias}}$ in radians, where $\sigma_\psi$ is expressed in radians \citep{1993A&A...274..968N,2001ApJ...553..341E}. The resulting single-pulse PPA distribution is shown in Figure~\ref{fig:profile_PA}(c).

To assess departures of the single-pulse PPA, $\psi$, from the integrated-profile PPA, $\psi_{\rm int}$, at the same pulse longitude, we calculated the $180^\circ$-periodic PPA difference as
\[
\Delta\psi=
\frac{1}{2}\operatorname{atan2}
\left[
\sin 2(\psi-\psi_{\rm int}),
\cos 2(\psi-\psi_{\rm int})
\right].
\]
The corresponding uncertainty-normalized residual was
\[
z=
\frac{\Delta\psi}
{\sqrt{\sigma_\psi^2+\sigma_{\psi,\rm int}^2}},
\]
where $\sigma_\psi$ and $\sigma_{\psi,\rm int}$ are the uncertainties of the single-pulse and integrated-profile PPAs,
respectively.
If the observed PPA spread were caused solely by measurement uncertainties and the high-S/N Gaussian approximation were applicable, $z$ would approximately follow a standard normal distribution.

As shown in Figure~\ref{fig:profile_PA}(c), the single-pulse PPAs in the MP exhibit a broad and structured distribution around the integrated-profile PPA. The uncertainty-normalized residuals in Figure~\ref{fig:profile_PA}(d) show that these deviations depend strongly on pulse longitude, with the most prominent structures occurring near $280^\circ$ and $310^\circ$, corresponding to the MP\_C2 and MP\_C3 regions, respectively. The persistence of these longitude-localized structures after normalization by the measurement uncertainties suggests that measurement uncertainty alone may not fully explain the observed PPA distribution. These structures also suggest component-dependent polarization behavior in the MP, with MP\_C3 exhibiting a different PPA distribution from that of the adjacent emission.
However, the present analysis does not uniquely locate its emission region or establish its magnetic-pole origin, because propagation effects and polarization-mode mixing may also contribute to the observed PPA distribution.


\begin{figure}
    \centering
    \includegraphics[width=3.5in,height=2in]{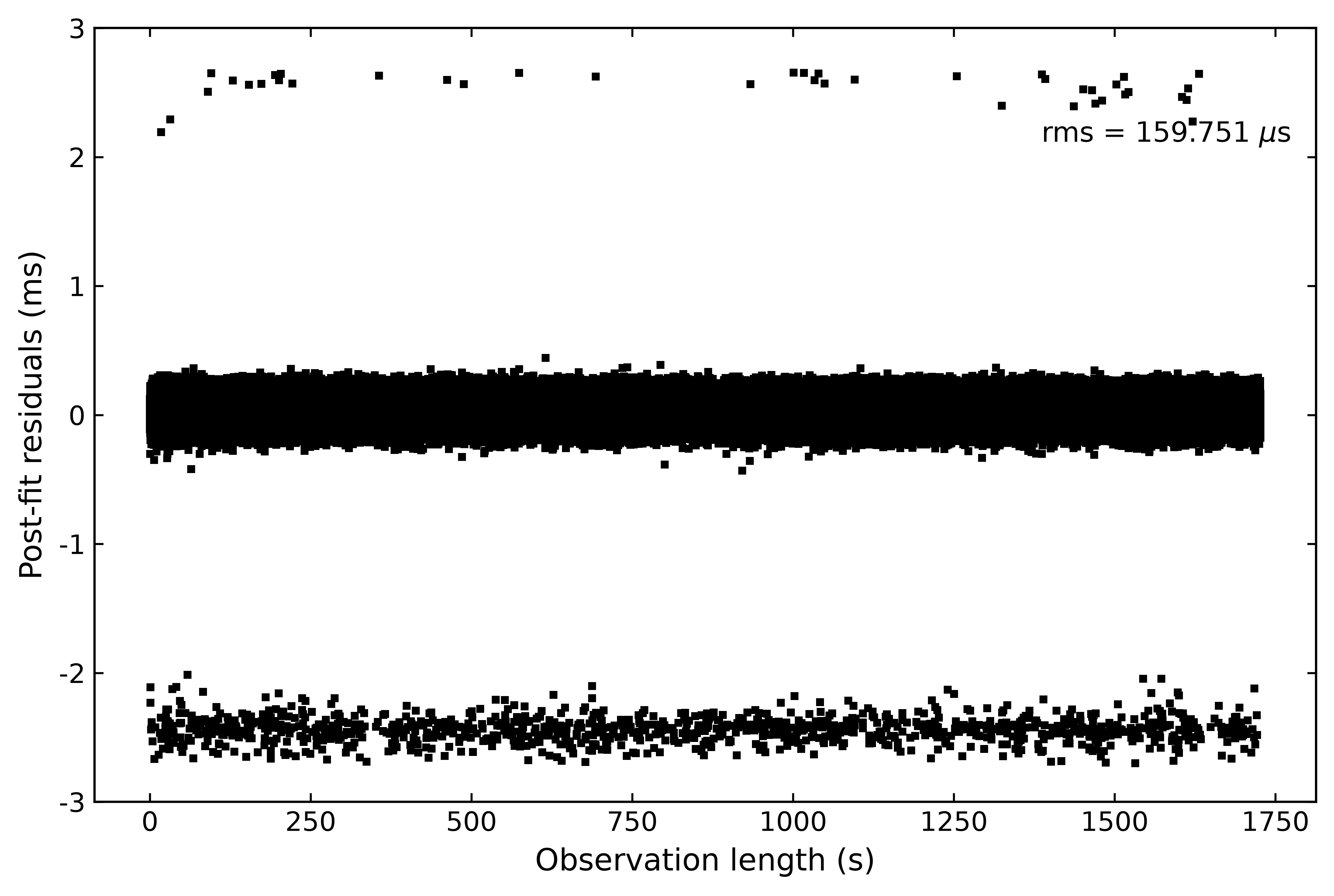}
\caption{
Post-fit timing residuals of PSR~J1857+0943.}
\label{fig:J1857_residuals}
\end{figure}

\begin{figure}
    \centering
    \includegraphics[width=3.5in]{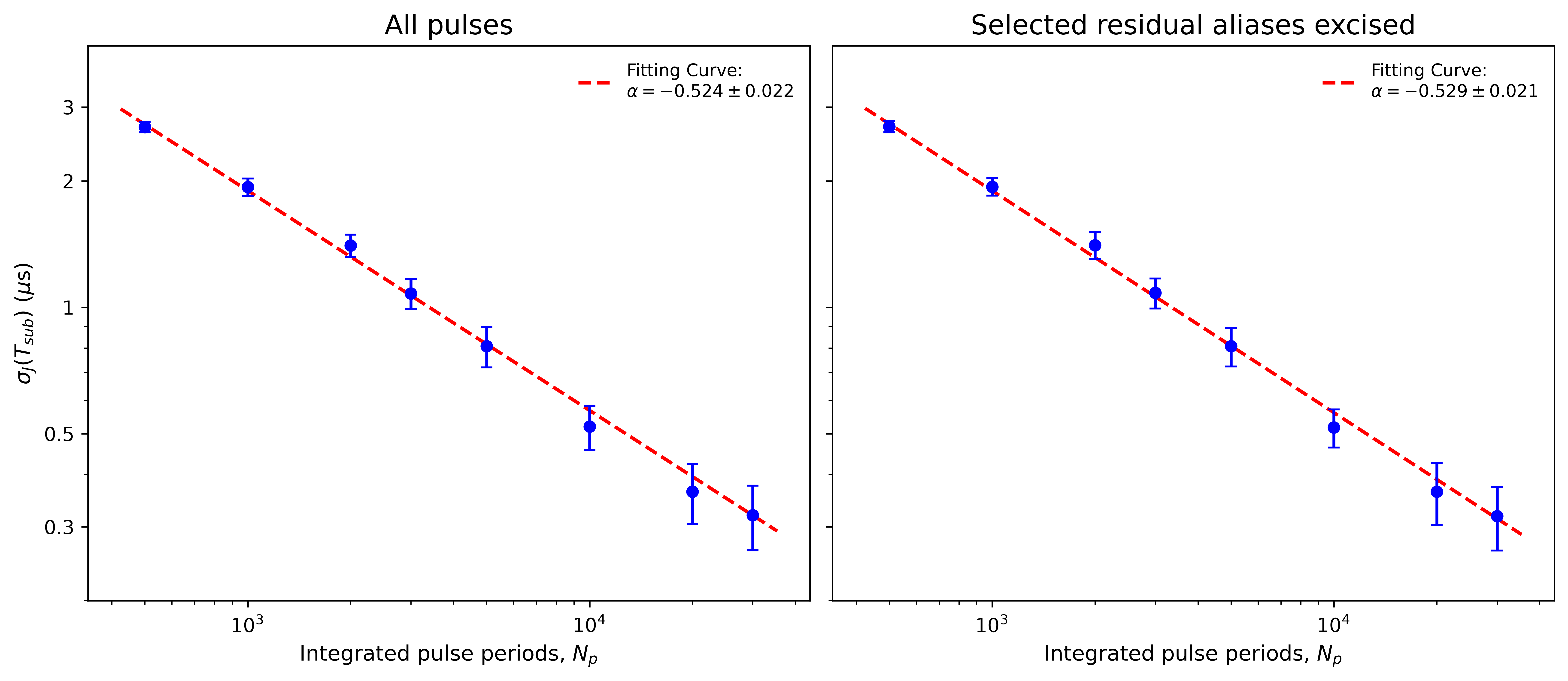}
    \caption{Jitter amplitude of PSR~J1857+0943 as a function of the number of integrated pulses, $N_{\rm p}$. The left panel shows the results obtained using all pulses, while the right panel shows the results after excluding the selected single-pulse residual aliases. The blue points show the measured $\sigma_J(N_{\rm p})$ values, and the red dashed curves show the weighted best-fitting power laws.}
    \label{fig:Jitter_VS_Tsub}
\end{figure}

\subsection{Single Pulse Timing and Jitter}

Pulse-to-pulse morphology variations, commonly referred to as jitter, have been extensively studied in pulsars employed for pulsar timing arrays (PTAs), as they impose a fundamental limit on the achievable timing precision \citep{2011MNRAS.418.1258O,2012MNRAS.420..361L,2014MNRAS.443.1463S,2019ApJ...872..193L,2021MNRAS.502..407P,2024ApJ...964....6W}.  
Both radiometer noise and jitter noise can contribute to short-timescale timing residuals, and both generally decrease approximately as the inverse square root of the number of integrated pulses. However, radiometer noise depends on the sensitivity of the observing system and the signal-to-noise ratio, whereas jitter is intrinsic to the pulsar and can become important in high-sensitivity observations.
The effect of jitter may be mitigated by selectively including an optimized subset of single pulses in the timing analysis \citep{2020ApJ...902L..13W}.

In our single-pulse timing analysis of PSR~J1857+0943, we identified two distinct emission components corresponding to the main pulse and the interpulse. 
When individual pulses were timed, the post-fit timing residuals exhibited a central residual band together with two displaced groups at approximately $\pm2.5$ ms (see Figure~\ref{fig:J1857_residuals}). The root-mean-square (rms) of the residuals for the total sample was measured to be $159.75~\mu$s. These separated clusters most likely arise when the template-matching procedure aligns single-pulse profiles dominated by the main pulse or interpulse with different components of the template, rather than directly indicating distinct emission modes. However, when the integration length was increased to include more than $\sim$1000 pulse periods per subintegration, this separation disappeared in PSR~J1857+0943. The merging of the residual distributions under longer averaging times suggests that this component-alignment ambiguity is effectively suppressed as both components become stably represented in the averaged profile, leading to a single stable mean profile.

We quantified the pulse jitter noise following the formalism of \citet{2021MNRAS.502..407P} and \citet{2024ApJ...964....6W}. 
Using frequency-averaged timing analysis, we divided the observation into contiguous subintegrations containing different numbers of pulses, $N_{\rm p}$. The corresponding integration time is $T_{\rm sub}=N_{\rm p}P$, where $P$ is the pulsar spin period. For each value of $N_{\rm p}$, ToAs were generated and the timing residuals were calculated relative to a timing model based on the PSRCAT ephemeris.
In this procedure, only the spin frequency was fitted, while all other timing parameters were held fixed, ensuring that the resulting residuals retained the short-timescale stochastic contributions intrinsic to the pulsar. The variance of these residuals therefore contains contributions from both radiometer noise and pulse-to-pulse jitter. Thus, the jitter amplitude, $\sigma_{J}(T_{\mathrm{sub}})$, was calculated as
\begin{equation}
   \sigma_{\mathrm{ToA}}^2(T_{\mathrm{sub}})=\sigma_{\mathrm{S/N}}^2(T_{\mathrm{sub}})+\sigma_{\mathrm{J}}^2(T_{\mathrm{sub}})  
\end{equation}
where $\sigma_{\mathrm{ToA}}^2(T_{\mathrm{sub}})$ represents the observed variance of the timing residuals, $\sigma_{\mathrm{S/N}}^2(T_{\mathrm{sub}})$ is the contribution due to radiometer noise from the receiver system, and $\sigma_{\mathrm{J}}^2(T_{\mathrm{sub}})$ represents the contribution from pulse jitter. 

To estimate $\sigma_{\mathrm{S/N}}^2(T_{\mathrm{sub}})$, we performed simulations using TEMPO2 plugins following the same method used in \citep{2021MNRAS.502..407P}. Ideal ToAs were first generated using the  $\mathtt{formIdeal}$ plugin based on the adopted timing model. Gaussian noise consistent with the radiometer-limited ToA uncertainties was then added using the $\mathtt{addGaussian}$ and $\mathtt{createRealisation}$ plugins. To robustly estimate the radiometer-noise variance, 1000 realisations were generated for every $T_{\mathrm{sub}}$. The mean variance from these simulations, which contains only the radiometer contribution and is diagonal by construction, was subtracted from the variance of the real FAST residuals. The resulting set of jitter variance allowed us to measure the evolution of the jitter noise amplitude as a function of sub-integration time. The uncertainty in the observed residual variance was estimated by bootstrap resampling of the observed residuals, while the uncertainty in the simulated radiometer-noise contribution was calculated from the standard error of the mean simulated variance. These uncertainties were propagated to the resulting jitter amplitude.

The measured jitter amplitude decreases from $2.74~\mu\mathrm{s}$ at $N_{\rm p}=500$ ($T_{\rm sub}=2.68$ s) to $0.34~\mu\mathrm{s}$ at $N_{\rm p}=30{,}000$ ($T_{\rm sub}=160.9$ s).
Figure~\ref{fig:Jitter_VS_Tsub} presents evolution of $\sigma_{J}(T_{\mathrm{sub}})$ as a function of $T_{\mathrm{sub}}$ in PSR\,J1857+0943. A weighted power-law fit of the form
\begin{equation}
\sigma_{J}(T_{\mathrm{sub}}) = A T_{\mathrm{sub}}^{\alpha}
\label{Eq:4}
\end{equation}
was applied, yielding $\alpha=-0.524\pm0.022$, consistent with the theoretical expectation of $\alpha = -0.5$ for uncorrelated pulse-to-pulse shape variations\citep[e.g.,][]{2010arXiv1010.3785C, 2014MNRAS.443.1463S,2012MNRAS.420..361L}. The fit was performed in logarithmic space using the relative jitter uncertainties as weights. To allow a direct comparison with previous studies, we fitted the evolution of jitter with integration time based on the theoretically expected ($T_{\mathrm{sub}}^{-1/2}$) scaling, and extrapolated to a one-hour integration, obtaining $\sigma_{J,1\mathrm{h}}=73.2\pm1.5~\mathrm{ns}$. 

To determine whether the displaced single-pulse residuals in Figure~\ref{fig:J1857_residuals} affect the inferred jitter level, we repeated the complete analysis after excluding the corresponding residual aliases. The timing residuals were first expressed in rotational phase and wrapped into the interval $[-0.5,0.5]$. This procedure identified 1,187 pulses, corresponding to 0.37\% of the 322,046-pulse sample. These pulses were assigned zero weight before the longer subintegrations were formed.
The same timing template, values of $N_{\rm p}$, timing fit, and radiometer-noise simulations were used for the complete and pulse-selected data sets. After the residual aliases were excluded, the free-index fit gives $\alpha=-0.529\pm0.021$, while the fixed-index fit gives $\sigma_{J,1{\rm h}}=73.0\pm1.6$ ns. This value differs from the result obtained using all pulses by only approximately 0.3\%, substantially less than the measurement uncertainty. The selected residual aliases therefore have no measurable effect on the inferred one-hour jitter amplitude.

\section{Discussion and Conclusion }\label{sec:disscussion}

\begin{figure}
    \centering
    \includegraphics[width=3.5in,height=2in]{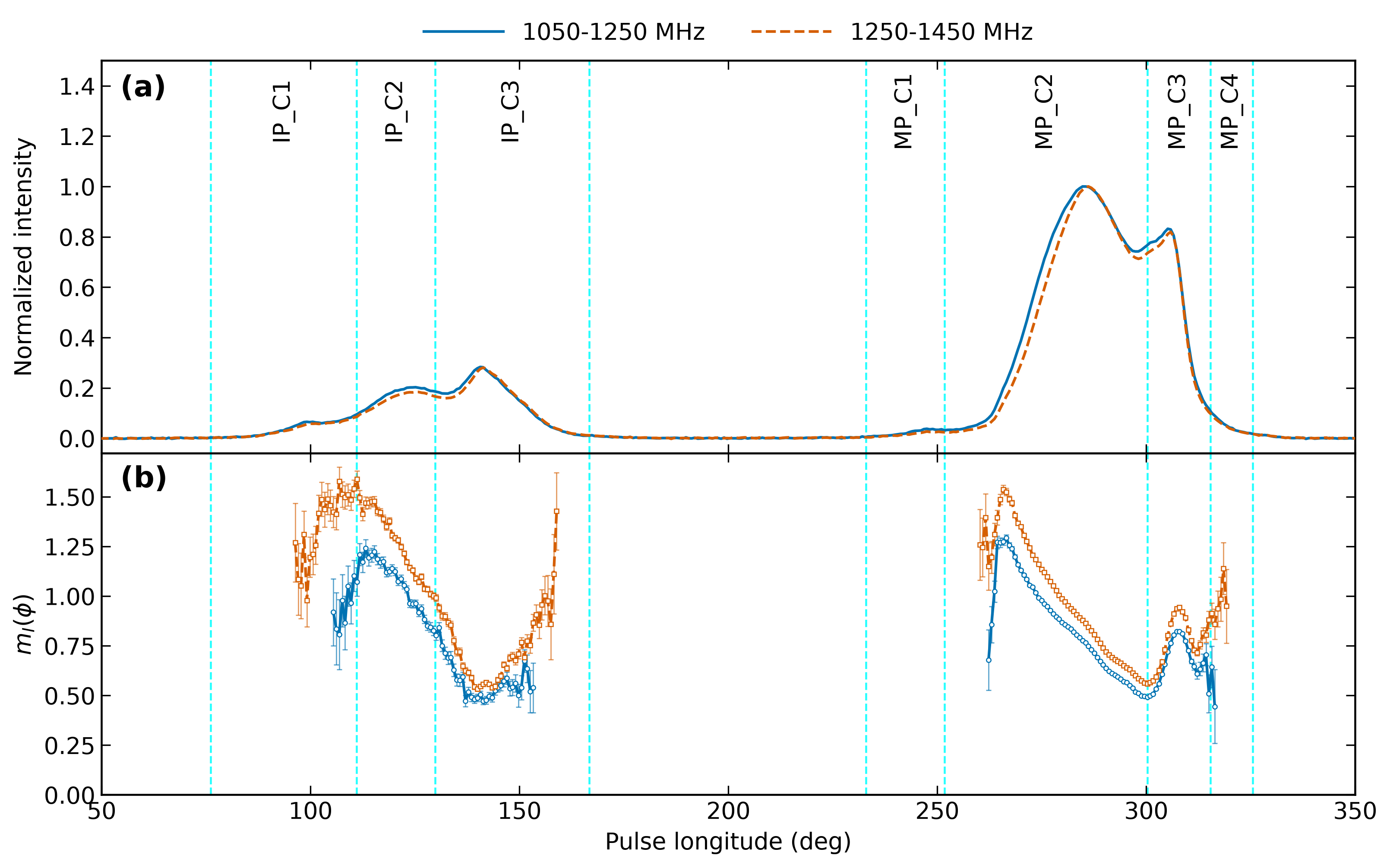}
\caption{Comparison of the average pulse profiles and longitude-resolved modulation indices of PSR J1857+0943 in the 1050--1250 MHz and 1250--1450 MHz sub-bands, shown in blue and orange, respectively. Panel (a) displays the average profiles, each normalized to its own peak intensity. Panel (b) shows the modulation index $m_I(\phi)$. The cyan dashed vertical lines mark the component boundaries derived from the Gaussian decomposition of the full-band (1050--1450 MHz) integrated profile.}
\label{fig:two_band_profile_modulation}
\end{figure}

We have investigated the single-pulse properties of  PSR~J1857$+$0943 using a 1726 s FAST observation comprising 322,046 individual rotations. The 2DFS confirms periodic amplitude modulation in MP\_C2, MP\_C3, IP\_C2, and IP\_C3. In all four components, the modulation feature is not spectrally resolved from $f_2=0$, and no robust evidence for systematic subpulse drifting is found. The corresponding modulation periods in IP\_C2 and IP\_C3 are approximately twice those in MP\_C2 and MP\_C3, although the physical mechanism responsible for this relationship remains unclear.

Periodic amplitude modulation has been reported in a number of MSPs \citep{2024ApJ...964....6W}. By contrast, subpulse drifting has been detected in only a few MSPs, with the reported drifting features generally being weak and diffuse rather than coherent \citep[e.g.,][]{2003A&A...407..273E,2016MNRAS.463.3239L,2023MNRAS.526.2156M}. In addition, since interpulse emission in MSPs is generally very weak, to our knowledge, periodic modulation in the interpulse of only one MSP, PSR~J1012+5307, has previously been
reported \citep{2024ApJ...964....6W}. PSR~J1857$+$0943 therefore becomes, to our knowledge, the second reported MSP in which periodic modulation is detected in the interpulse region. These markedly different modulation behaviors among the profile components of PSR~J1857$+$0943 demonstrate that the components undergo distinct fluctuation processes.

The LRCCF provides complementary information about the component-dependent behavior revealed by the 2DFS. The nearly vertical correlation structure in MP\_C3 is consistent with amplitude modulation without a clear systematic phase drift. The full-band lag-one LRCCF shows localized anti-correlations between MP\_C3 and the IP components IP\_C1 and IP\_C2. Under the lag convention adopted here, enhanced MP\_C3 emission in pulse $n$ tends, on average, to be followed by reduced IP emission in pulse $n+1$. The retention of the negative MP\_C3--IP correlations in both sub-bands suggests that these features are comparatively stable in sign across the observed frequency range. 
By contrast, the correlation between MP\_C3 in pulse $n$ and part of MP\_C2 in pulse $n+1$ reverses sign between the two sub-bands, showing a stronger frequency dependence than the MP\_C3--IP anticorrelations.

The absence of comparable coherent structures in the pulse-order-randomized maps further shows that the observed LRCCF features depend on the original pairing between successive pulses. The negative MP\_C3--IP correlations persist in both sub-bands. Although small instrumental contributions cannot be completely excluded, these results support a possible physical coupling between the corresponding longitude ranges.

Figure~\ref{fig:two_band_profile_modulation} shows the average profiles and longitude-resolved modulation indices in the 1050--1250 and 1250--1450 MHz sub-bands. The normalized MP and IP profiles are broadly similar, with small differences in the relative component intensities. Where the modulation index can be reliably measured in both sub-bands, it decreases toward the MP\_C2--MP\_C3 boundary and rises within MP\_C3. This trend is present in both sub-bands, although the modulation-index values differ. Near the weak profile edges, the measurements are more sensitive to the noise correction and should be compared with caution. The similar modulation-index trends do not imply similar correlations between successive pulses. The modulation index measures the strength of intensity fluctuations relative to the mean at each longitude, whereas the lag-one LRCCF measures how intensities at two longitudes vary together across successive pulses. The MP\_C3--MP\_C2 correlation can therefore reverse sign without a comparable change in the average profile or modulation index. Changes in the relative contributions of overlapping components could contribute to this reversal, but the present comparison does not establish that explanation.

The localized lag-one anti-correlations between MP\_C3 and the interpulse components IP\_C1 and IP\_C2 may reflect a physical association between the corresponding pulse-longitude ranges. One possible interpretation is that, although MP\_C2 and MP\_C3 are adjacent in the average profile, they do not share the same physical origin, and that MP\_C3 is instead associated with the same magnetic pole as the IP. In a conventional low-altitude, axisymmetric conal-beam model, assigning MP\_C3 and the IP to the same magnetic pole is geometrically difficult. This is not necessarily the case for patchy or fan beams. In these models, widely separated pulse windows can arise from distinct active regions or magnetic flux tubes associated with the same pole, without requiring continuous emission over more than $180^\circ$ of pulse longitude \citep{1988MNRAS.234..477L,2014ApJ...789...73W,2019MNRAS.490.4565J}. Although our data do not distinguish among the possible beam geometries, the longitude separation alone does not exclude a common-pole origin for MP\_C3 and the IP. 
MP\_C4 immediately follows MP\_C3 in pulse longitude and may represent a weaker extension of the same emission region. However, its substantially lower intensity prevents a reliable characterization of its fluctuation and LRCCF properties. The absence of a comparable detected feature in MP\_C4 therefore cannot be regarded as evidence against this possibility. Longer or higher-sensitivity single-pulse observations are required to determine whether MP\_C4 participates in the proposed MP\_C3--IP association.

\citet{2017MNRAS.472.4598D} showed that coherent mode addition and propagation can produce broad, loop- or V-shaped PPA bifurcations that are accompanied by enhanced $|V|$ and reduced $L/I$ at the transition longitudes. For PSR~J1857$+$0943, the single-pulse PPA distributions near pulse longitudes of $280^\circ$ and $310^\circ$ are compact and localized rather than broad or looped, they are not accompanied by strong $|V|$ peaks or $L/I$ minima and occur in regions with component-dependent fluctuation behavior. These features deviate significantly from the average-profile PPA curve and do not closely resemble the polarization-angle bifurcations described by \citet{2017MNRAS.472.4598D}. They suggest that the emission in MP\_C3 may include a contribution from an emission region distinct
from those of the adjacent MP components, although propagation effects and polarization-mode mixing cannot be fully ruled out.

Taken together, the observation that the modulation periods of the IP components are approximately twice that of MP\_C3, the longitude-resolved modulation-index structure within MP\_C3, its localized lag-one LRCCF anti-correlations with IP\_C1 and IP\_C2, and its localized single-pulse PPA features indicate that MP\_C3 exhibits component-dependent intensity and polarization properties within the MP. These combined properties lend support to a possible cross-pole interpretation of the trailing MP emission, in which MP\_C3 and possibly the much weaker MP\_C4 are associated with the same magnetic pole as the IP. However, broader magnetospheric coupling between the MP\_C3 and IP longitude ranges remains a viable alternative explanation.

The jitter measurements of PSR~J1857+0943 follow the $T_{\mathrm{sub}}^{-1/2}$ scaling expected for effectively uncorrelated pulse-to-pulse variations \citep{2010arXiv1010.3785C,2012MNRAS.420..361L,2014MNRAS.443.1463S}. Since radiometer noise follows the same integration-time dependence, the jitter amplitude in our analysis is determined by subtracting the simulated radiometer-noise variance from the observed residual variance, rather than from the scaling relation alone. The one-hour jitter amplitude inferred from our analysis is consistent with the previous measurement of PSR~J1857+0943 \citep{2023MNRAS.520.2747P}, supporting the robustness of the measured intrinsic timing-noise level.

The negligible change in the inferred jitter level after excluding the displaced single-pulse residual aliases indicates that these rare template-alignment ambiguities are effectively averaged out when hundreds or more pulses are integrated. Simple selection based solely on single-pulse timing residuals therefore provides no measurable improvement in the timing precision for this observation. This result does not rule out improvements from methods that retain more detailed pulse-shape information, such as profile-based classification or time tagging \citep{2015MNRAS.452..607K}, but testing such methods would require a dedicated analysis beyond the scope of this work.

\begin{acknowledgments}
This work was supported by the Strategic Priority Research Program of the Chinese Academy of Sciences, Grant No.XDA0350501, the Major Science and Technology Program of Xinjiang Uygur Autonomous Region (No.2022A03013-4), the Guizhou Provincial Basic Research Program (Natural Science) (No. Qiankehejichu-MS(2025)263), the Guizhou Provincial Science and Technology Projects (Nos. QKHFQ[2023]003, QKHPTRC-ZDSYS[2023]003, QKHFQ[2024]001-1), Guizhou Provincial Scientific and Technological Program The Postdoctoral Research Station of the Guizhou Radio Astronomical Observatory. (No.QKHPTRC[2021]Postdoctoral Research Station-001), the National Natural Science Foundation of China (No. 12603070, 12041304), the Project funded by China Postdoctoral Science Foundation(No. 2023M743517). This work made use of the data from the Five-hundred-meter Aperture Spherical Radio telescope, which is a Chinese national mega-science facility, operated by the National Astronomical Observatories, Chinese Academy of Sciences.

\end{acknowledgments}

\vspace{5mm}







\bibliography{sample631}{}
\bibliographystyle{aasjournal}


\appendix
\section{Derivation of the Modulation Index}\label{App:Appendux_A}
In the case of using a single off-pulse window, we first consider the statistical properties of the off-pulse region. Let $N$ denote the number of phase points in the off-pulse region, where each point follows a normal distribution with mean $\mu_0$ and standard deviation $\sigma_0$. During the baseline removal process, the mean of the off-pulse region is calculated for each pulse, and the baseline is subtracted. This mean follows a normal distribution with mean $\mu_0$ and standard deviation $\frac{\sigma_0}{\sqrt{N}} )$.

After baseline removal, the off-pulse mean becomes zero, and its standard deviation is reduced to 
\begin{equation}
\sigma_1 = \sqrt{\frac{N-1}{N}} \sigma_0 
\end{equation}

For the on-pulse region, the observed standard deviation $\sigma_{\text{on}}$ and mean $\mu_{\text{on}}$ relate to the intrinsic pulse properties through
\begin{equation}
\sigma_{\text{on}} = \sqrt{\sigma_{\text{pulse}}^2 + \frac{N+1}{N} \sigma_0^2}, \quad
\mu_{\text{on}} = \mu_{\text{pulse}},
\end{equation}
where $\sigma_{\text{pulse}}$ and $\mu_{\text{pulse}}$ are the true standard deviation and mean of the pulse emission.
Substituting the expressions for the baseline variance, we obtain
\begin{equation}
\sigma_{\text{pulse}} = \sqrt{\sigma_{\text{on}}^2 - \frac{N+1}{N-1} \sigma_1^2},
\end{equation}
and thus, the corrected modulation index is
\begin{equation}
m = \frac{\sigma_{\text{pulse}}}{\mu_{\text{on}}} = \frac{\sqrt{\sigma_{\text{on}}^2 - \frac{N+1}{N-1} \sigma_1^2}}{\mu_{\text{on}}}.
\end{equation}
The above derivation indicates that baseline subtraction introduces a systematic effect on the estimation of the modulation index.
When the standard deviation of the off-pulse region, denoted as $\sigma_1$, is used as the reference noise level, the pulse-phase variance $\sigma_{\text{pulse}}^2$ is effectively overestimated by $\frac{2}{N}\sigma_0^2$. For most pulsars, $\sigma_0$ is typically comparable to or even larger than $\sigma_{\text{pulse}}$, leading to an overestimation of the modulation index, especially in pulse-phase bins near the profile edges where the mean intensity is relatively low.
To preserve the variance characteristics of different phase bins after baseline removal, the corresponding regions should be multiplied by a correction factor of $\frac{N+1}{N-1}$. This adjustment ensures that the variance properties remain consistent across all pulse-phase bins. Moreover, the baseline subtraction process itself reduces the single-pulse signal-to-noise ratio along the pulse sequence by a factor of $\frac{N+1}{N}$. Therefore, the baseline should ideally be estimated using as many phase bins as possible to minimize this effect.

The uncertainty of the corrected modulation index was estimated through standard error propagation as
\begin{equation}
\delta m =
\left\{
\left[
\frac{\sigma_{\rm on}}
{\mu_{\rm on}\sqrt{\sigma_{\rm on}^2-K\sigma_1^2}}
\delta\sigma_{\rm on}
\right]^2
+
\left[
\frac{K\sigma_1}
{\mu_{\rm on}\sqrt{\sigma_{\rm on}^2-K\sigma_1^2}}
\delta\sigma_1
\right]^2
+
\left[
\frac{m}{\mu_{\rm on}}
\delta\mu_{\rm on}
\right]^2
\right\}^{1/2},
\end{equation}
where $K=(N+1)/(N-1)$.
This expression shows that the uncertainty of $m$ is mainly controlled by two factors: the accuracy of the variance correction and the significance of the mean pulse intensity. In particular, when the corrected variance,
\begin{equation}
\sigma_{\rm on}^2-K\sigma_1^2,
\end{equation}
is small, the first two terms in the error propagation become large, making the derived modulation index highly uncertain. This situation commonly occurs near the edges of the average pulse profile, where the mean intensity is weak and the noise correction can be comparable to the measured variance. Therefore, modulation indices obtained from low-significance phase bins should be interpreted with caution.



\end{document}